\documentclass[fleqn,usenatbib]{mnras}
\usepackage{newtxtext,newtxmath}
\usepackage[T1]{fontenc}

\DeclareRobustCommand{\VAN}[3]{#2}
\let\VANthebibliography\thebibliography
\def\thebibliography{\DeclareRobustCommand{\VAN}[3]{##3}\VANthebibliography}

\usepackage{graphicx}	
\usepackage{amsmath}	
\usepackage{xspace}
\usepackage{xcolor}     
\usepackage{soul}

\newcommand{\Nsample}{$22$\xspace}

\newcommand{\atlas}{$\mathrm{ATLAS}^{\mathrm{3D}}$\xspace}
\newcommand{\re}{\ensuremath{R_{\mathrm{e}}}\xspace}
\newcommand{\Msun}{\ensuremath{\mathrm{M}_\odot}\xspace}
\newcommand{\Lsun}{\ensuremath{\mathrm{L}_\odot}\xspace}

\newcommand{\kms}{\ensuremath{\mathrm{km\,s}^{-1}}\xspace}
\newcommand{\tr}{\ensuremath{\mathrm{T}_{R_\mathrm{e}} =  (1 - p^2)/(1 - q^2)}\xspace}
\newcommand{\Tr}{\ensuremath{\mathrm{T}_{R_\mathrm{e}}\xspace}}

\title[Evidence against the Bulge-Halo Conspiracy]{The MAGPI Survey: Evidence Against the Bulge-Halo Conspiracy}

\author[Derkenne et al.]{
C. Derkenne,$^{1,2}$
R. M. McDermid,$^{1,2}$\thanks{E-mail: richard.mcdermid@mq.edu.au}
G. Santucci,$^{3,2}$
A. Poci,$^{4}$
S. Thater,$^{5}$
S. Bellstedt,$^{3}$
J. T. Mendel,$^{2,6}$
\newauthor
C. Foster,$^{2,7}$
K. E. Harborne,$^{3,2}$
C. D. P. Lagos,$^{2,3}$
E. Wisnioski,$^{2,6}$
S. Croom,$^{2,8}$
R-S. Remus,$^{9}$ 
\newauthor
L. M. Valenzuela,$^{9}$
J. van de Sande,$^{2,8}$
S. M. Sweet,$^{2,10}$ 
and B. Ziegler$^{5}$
\\
$^{1}${Research Centre for Astronomy, Astrophysics, and Astrophotonics, School of Mathematical and Physical Sciences, Macquarie University,NSW 2109, Australia}\\
$^{2}${ARC Centre of Excellence for All Sky Astrophysics in 3 Dimensions (ASTRO 3D), Australia}\\
$^{3}${International Centre for Radio Astronomy Research, The University of Western Australia, 35 Stirling Highway, Crawley, WA 6009, Australia}\\
$^{4}${Centre for Extragalactic Astronomy, University of Durham, Stockton Road, Durham DH1 3LE, United Kingdom}\\
$^{5}${Department of Astrophysics, University of Vienna, Türkenschanzstrasse 17, 1180 Vienna, Austria}\\
$^{6}${Research School of Astronomy and Astrophysics, Australian National University, Canberra, ACT 2611, Australia}\\
$^{7}${School of Physics, University of New South Wales, Sydney, NSW 2052, Australia}\\
$^{8}$Sydney Institute for Astronomy, School of Physics, University of Sydney, NSW 2006, Australia\\
$^{9}$Universit\"{a}ts-Sternwarte, Fakult\"{a}t f\"{u}r Physik, Ludwig-Maximilians-Universit\"{a}t M\"{u}nchen, Scheinerstr. 1, 81679 M\"{u}nchen, Germany\\
$^{10}$School of Mathematics and Physics, University of Queensland, Brisbane, QLD 4072, Australia\\
}

\date{Accepted XXX. Received YYY; in original form ZZZ}

\pubyear{2024}

\begin{document}
\label{firstpage}
\pagerange{\pageref{firstpage}--\pageref{lastpage}}
\maketitle

\begin{abstract}
Studies of the internal mass structure of galaxies have observed a `conspiracy' between the dark matter and stellar components, with total (stars $+$ dark) density profiles showing remarkable regularity and low intrinsic scatter across various samples of galaxies at different redshifts. Such homogeneity suggests the dark and stellar components must somehow compensate for each other in order to produce such regular mass structures. We test the conspiracy using a sample of \Nsample galaxies from the `Middle Ages Galaxy Properties with Integral field spectroscopy' (MAGPI) Survey that targets massive galaxies at $ z \sim 0.3$. We use resolved, 2D stellar kinematics with the Schwarzschild orbit-based modelling technique to recover intrinsic mass structures, shapes, and dark matter fractions. This work is the first implementation of the Schwarzschild modelling method on a sample of galaxies at a cosmologically significant redshift. We find that the variability of structure for combined mass (baryonic and dark) density profiles is greater than that of the stellar components alone. Furthermore, we find no significant correlation between enclosed dark matter fractions at the half-light radius and the stellar mass density structure. Rather, the total density profile slope, $\gamma_{\mathrm{tot}}$, strongly correlates with the dark matter fraction within the  half-light radius, as $\gamma_{\mathrm{tot}} = (1.3 \pm 0.2) f_{\mathrm{DM}} - (2.44 \pm 0.04)$. Our results refute the bulge-halo conspiracy and suggest that stochastic processes dominate in the assembly of structure for massive galaxies. 
\end{abstract}




\section{Introduction}
\label{sec:introduction}

Since the discovery that the observed kinematics of galaxies do not follow the gravitational potential established by the stellar mass alone, dark matter has been a ubiquitous ingredient in the field of galaxy evolution, and is thought to dominate the Universe's mass budget \citep{blumenthal_1984}. The theorised assembly history of galaxies has baryonic matter fall into a potential well seeded by dark matter in the early Universe \citep{white_and_rees,wechsler_2018_connection}. As such, the inclusion of dark matter in the gravitational potential of galaxy dynamical models is now standard. Of particular interest is how dark matter interfaces with the baryonic matter of galaxies, for example the scale at which it becomes the dominant mass component, and what impact it has on the global potential and mass distribution within a galaxy. 

Efforts to measure the combined baryonic and dark matter content of galaxies have often assumed the distribution for the total mass density profile, where the  combination of the baryonic and dark matter profiles is approximated as a simple power law, of form $ \rho \propto r^{\gamma}$, with $\rho$ the mass density, $r$ the galactocentric radius, and $\gamma$ the slope of the profile (here we adopt the convention of $\gamma < 0$). Measurements of this slope using dynamical studies \citep{cappellari_2015_small,poci_2017_syst,bellstedt_sluggs_2018,li_manga_2019,versic_2024}, lensing \citep{etherington_2023_beyond}, and combinations of the two \citep{auger_sloan_2010,barnabe_two_2011,ruff_sl2s_2011,bolton_2012_BOSS,sonn_sl2s_2013}, have found galaxies tend towards mass density profiles with near-isothermal ($\gamma \sim -2$) slopes and limited intrinsic scatter, despite a significant range of stellar luminosity profiles. This observation has been dubbed the ‘bulge-halo conspiracy', as the stellar component and dark matter component must jointly `conspire' to yield isothermal total profiles with low scatter across (mainly early-type) galaxy populations \citep{dutton_2014_bulge}. The conspiracy indicates remarkable homogeneity of mass structures for galaxies from the local Universe to $z\sim 1$. 

The physical origins of such a conspiracy are unclear. Although both components (baryonic and dark) must exist within the same gravitational potential, the stellar mass structure of galaxies is highly variable across galaxy types. This variability is reflected by a  non-universal initial mass functions (IMF) both within \citep{McConnell_2016,Conroy_2017,dokkum_2017_IMF,navarro_2019_fornax} and between galaxies \citep{cappellari_2012_IMF,smith_2020_evidence,martin_navarro_2021,thater_2023}, and measurements of galaxy stellar structure such as the S\'{e}rsic index, which show a large range in values \citep{Graham2013,lange_2015_GAMA,Zahid_2017}. The combination of the stellar component and an independent dark matter component should, naively, increase the observed variance of a population of total density profiles, instead of the small scatter and near isothermal profiles observed.

The conspiracy was explicitly tested by measuring  de-coupled stellar and dark matter profiles of \atlas galaxies with Jeans \citep{cappellari_measuring_2008,cappellari_2020} dynamical modelling \citep{poci_2017_syst}. \citet{poci_2017_syst} found that the intrinsic scatter in the slope of the combined density profiles was only marginally larger than that of the stellar profiles alone, and could therefore make no strong assessments about the validity of the `conspiracy'. Their approach included two distinct models; one in which the gravitational potential was treated a single ``total-mass'' density profile, and another in which the stars and dark matter were modeled explicitly. Jeans modelling method used assumes galaxies have an axisymmetric potential, and \citet{poci_2017_syst} further assumed a cylindrically aligned velocity ellipsoid and a global orbital anisotropy term, which might not reflect the true orbital structures present in galaxies. Crucially, the majority of the \atlas sample has kinematic data only interior to the half-light radius scale, constraining the stellar and dark components on a small physical region. Within the (presumably baryon dominated) half-light radius, it may be expected that the total and stellar mass structures closely trace each other.

In this work, we leverage `Middle Ages Galaxy Properties with Integral field spectroscopy' (MAGPI) Survey\footnote{Based on observations obtained at the Very Large Telescope (VLT) of the European Southern Observatory (ESO), Paranal, Chile (ESO program ID 1104.B-0536)} data to construct Schwarzschild orbit-based models \citep{schwarzschild_1979_numerical} of massive galaxies at $ z \sim 0.3$. The benefit of studying this epoch is that the data frequently extends to two or even three half-light radii with a resolution that is comparable to local surveys such as the Sydney Australian Astronomical Observatory Multi-object Integral Field Spectrograph (SAMI) Survey \citep{croom_2012_SAMI,Santucci_2022}. Furthermore, it is valuable to investigate the dynamical state of galaxies during this period of the Universe's history. We present triaxial Schwarzschild dynamical models of \Nsample MAGPI galaxies, constructed using the {\sc{dynamite}} code \citep{jethwa_2020_dynamite,thater_2022_mirroring}, which relaxes some of the assumptions necessary for Jeans dynamical modelling. As an orbit-based technique, Schwarzschild models that accurately reconstruct galaxy observables (flux and higher-order stellar kinematics) provide insight into the intrinsic properties of galaxies, such as 3D shape, orbital structure, and dark matter fractions. We use these detailed  models to investigate how the internal mass of a galaxy is distributed between its stellar and dark components, and test the predictions of the bulge-halo conspiracy. 

In Section~\ref{sec:data} we introduce the MAGPI data, and present how the inputs to the Schwarzschild models are derived in Section~\ref{sec:methods}. The construction of the Schwarzschild models is given in Section~\ref{sec:methods:schwarzshild_models}, with our results from these models presented in Section~\ref{sec:results}. We discuss our results in Section~\ref{sec:discussion} and summarise this work in Section~\ref{sec:conclusion}.

Throughout this article, we adopt a flat $\Lambda$CDM cosmology with $H_0 = 67.1\ \kms \mathrm{Mpc}^{-1}$ and $\Omega_m = 0.3175$, corresponding to the Planck 2013 cosmology \citep{planck_collab_2013}. All scales are converted (angular and physical units) using the angular diameter distance given by the object's redshift and the above cosmology. 

\section{Data Sample}
\label{sec:data}

 In this work we use the 30 objects for which Jeans dynamical models were constructed by \citet{derkenne_2023_MAGPI}, although due to sample cuts described in Section~\ref{sec:methods:parameter_space}, we present results for only 22 of these objects. Modelling the same sample with two different techniques provides a valuable comparison point, presented in Section~\ref{sec:results:jeans}. Importantly, in \citet{derkenne_2023_MAGPI} we showed that the spatial resolution of this sample is high enough that the recovered mass distribution of the models are not biased due to point-spread function (PSF) effects, which is important to consider when constructing dynamical models of objects at any appreciable redshift. The data are deep enough to provide up to $\sim 4$ half-light radii of stellar kinematic coverage in some cases.  
 
 The sample is  drawn from the MAGPI Survey, a VLT/MUSE Large Program which targets massive galaxies in the Universe's `middle ages'. The survey uses MUSE in wide-field mode, yielding 60 arcsecond on a side fields. Resulting spectra span the range $4650 - 9300\,$\AA\ in steps of $1.25\,$\AA\ per pixel, with ground-layer adaptive optics typically resulting in PSF full-width half-maxima (FWHM) of $\sim 0.6$ arcsec, equivalent to 4.629 kpc at $z = 0.3$. Fields are observed for a total on-source integration time of 4.4 hours per field. The detailed aims of the MAGPI Survey are presented in \citet{MAGPI}. Although the survey officially targets 60 massive primary galaxies, the fields obtained to date also contain many secondary objects that have adequate angular resolution for resolved studies.
 
 The full details of the data reduction process will be presented by Mendel et. al. (in prep.), but we provide an overview  here. Data reduction is based on the MUSE reduction pipeline \citep{weilbacher_2020_data}. The Zurich Atmosphere Purge ({\sc{zap}}) software is used for sky subtraction \citep{soto_2016_zap}. Segmentation maps for all sources are made using {\sc{profound}} \citep{robotham_2018_profound}. A `minicube' is constructed for each source identified from the segmentation maps for every field, with the identified source at the centre and minicube extent such it encompasses the full dilated segment. Images for each field are made from the collapsed spectral cube of each field. 

 Integrated stellar masses for the sample are calculated using the spectral energy distribution fitting software {\sc{prospect}} \citep{robotham_2020_prospect}, assuming a \cite{chabrier_2002_galactic} initial mass function. The spectral energy distribution is fit using pixel-matched imaging in the \textit{ugriZYJHKs} bands from the GAMA Survey \citep{bellstedt_2020_gama}, with photometry derived by the MUSE-based {\sc{profound}} segmentation maps for each galaxy. The sample spans stellar masses between $\log_{10}(M_{\star}/\Msun) = 10.4$ and $\log_{10}(M_{\star}/\Msun) = 11.6$, with a sample mean redshift of $z = 0.304$. From team visual morphology classifications, our sample is dominated by lenticular and early-type morphologies, with only 4/\Nsample objects showing evidence of spiral arms (Foster et. al., in prep).  

\begin{figure}
\centering
\includegraphics[width=0.8\linewidth]{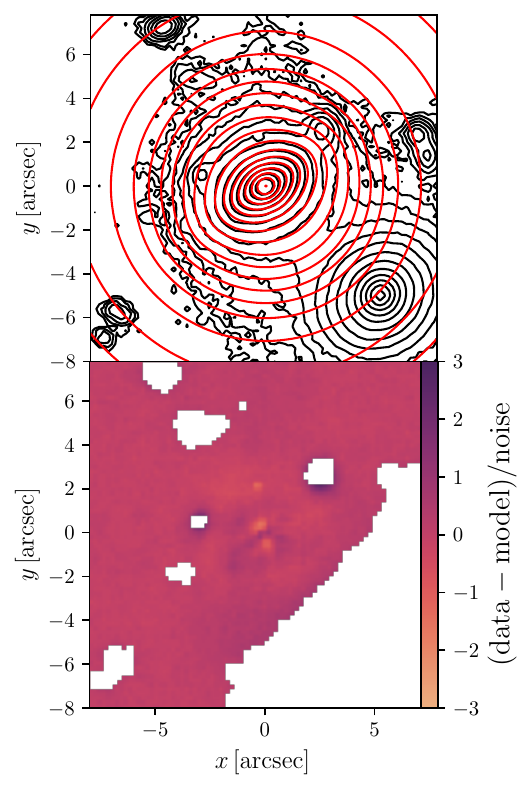}
\caption{The surface brightness model for MAGPI object 1525170222. The top panel shows the isophotes of the galaxy in black and of the MGE model overlaid in red, in steps of 0.5 $\mathrm{mag}/\mathrm{arcsec}^{2}$. Neighbouring objects are shown in the plot, but masked from the actual fit. The bottom panel shows the residuals of the model as (data-model) divided by the noise of the flux image. White regions indicate where neighbours were masked based on the {\sc{profound}} segmentation maps.}
\label{fig:mge}
\end{figure}

\section{Inputs to Schwarzschild models}
\label{sec:methods}
The two basic requirements when constructing a Schwarzschild model are 2D stellar kinematics with as many higher-order moments as feasible (dependent on the quality of the spectral data), and a description of the galaxy surface brightness. In the following sections we describe our modelling of the galaxy surface brightness and our extraction of the 2D stellar kinematics. Additionally, for our modelling approach, we transform the surface brightness models to stellar mass models via spectral measurements of stellar mass-to-light ratios.

\subsection{Galaxy surface brightness models}
\label{sec:methods:MGES}
We parameterise the galaxy light as a series of $N$ concentric 2D Gaussians \citep{emsellem_1994_multi} using the Python package {\sc{mgefit}} \citep{MGE}. This  multi-Gaussian expansion (MGE) is a projected model of the surface brightness, which is used to describe the intrinsic (deprojected) luminosity density of the stellar tracer within a given gravitational potential. The fitting process is identical to that used by \citet{derkenne_2023_MAGPI}, where we presented detailed explorations of how pixel sampling and resolution effects impact MGE models, benchmarked against Hubble Space Telescope results. We use a constant position angle in the MGE fits, as a radially varying position angle was not warranted by the fit quality of the models compared to the data.
We choose the $r$-band images built from the MUSE data cubes for their depth, with the initial image size selected to have sides of 20 \re from the {\sc{profound}} estimate. The choice of (observed-frame) $r$-band ensures that the observed frame light used to create the surface brightness models is similar to the spectral window used to measure the galaxy kinematics. Due to the redshift of the targets, the observed light in the $r$-band corresponds  approximately to rest-frame $g$-band of the sources. Therefore, in our analysis, we calibrate the projected surface luminosity density of each source in the (rest-frame) $g$-band (Eq. \ref{eq:surface_dens}). All neighbouring sources are masked out using the segmentation maps. MGEs are also made of stacked point sources in each field to model the point-spread function (PSF) in order to analytically convolve the surface brightness model with the PSF before comparing to observations. In this way, the best-fit model provides a deconvolved description of the surface brightness, corrected for the effects of PSF.

We require the galaxy surface brightness models to be in units of $\Lsun/\mathrm{pc}^2$ (and eventually the mass models in $\Msun/\mathrm{pc}^2$). First, the enclosed counts ($C$) of each Gaussian component are re-normalised to peak surface brightness (in counts per pixel), $C_0$: 
\begin{equation}
\begin{aligned}
\label{eq:gauss_height}
    C_0 = \frac{C}{2\pi\sigma^2q},
\end{aligned}
\end{equation}
with $q$ the axial ratio of the Gaussian and $\sigma$ the dispersion. The counts are converted to ST magnitudes ($m_{\mathrm{ST}}$) via
\begin{equation}
\begin{aligned}
\label{eq:ST}
    m_{\mathrm{ST}} = 28.9 - 2.5\log_{10}C_0.
\end{aligned}
\end{equation}
The $g$-band peak surface brightness in magnitudes per square arcsecond is given by 

\begin{equation}
\begin{aligned}
\label{eq:surf_brightness}
     \mu = m_{\mathrm{ST}} + 2.5\log_{10}(s)^2  - A_r + k - 2.5\log_{10}(1+z)^{5},
\end{aligned}
\end{equation}
where $s$ is the pixel scale, $A_r$ is the line-of-sight dust correction from \citet{schlafly_2011}, and $k$ is the k-correction which also converts from $r$ to $g$ band. The k-correction is necessary as the MAGPI sources are sampled at various rest-frame frequencies due to their redshift, which makes comparing across sources without a correction invalid. We follow \citet{hogg_2002} equation 8 to perform the correction, and in practise calculate the difference in magnitudes between the observed frame spectrum and its rest frame counterpart in the $r$ and $g$ bands, respectively. The final term in Eq~\ref{eq:surf_brightness} is the cosmological surface brightness dimming correction, following the derivation given by \citet{Whitney_2020}  for ST magnitudes. We convert the $g$-band surface brightness to a projected 2D surface density, $I'$, in units of $L_{\odot}/\mathrm{pc}^2$,  with:
\begin{equation}
\begin{aligned}
\label{eq:surface_dens}
 I' = \left(\frac{64800}{\pi}\right)^2 10^{0.4(M_g - \mu)},
\end{aligned}
\end{equation}
where $M_g$ is the absolute magnitude of the sun in the $g$-band from \citet{Willmer_2018}. Finally, the dispersion of each Gaussian component is transformed from units of pixels to arcseconds by multiplying by the MUSE pixel scale. We show an example of an MGE fit and its residuals in Figure~\ref{fig:mge}. In the subsequent analysis, we define the half-light radius, \re, as the semi-major axis of the best-fit MGE model ellipse which contains half the model light.

\subsection{Stellar kinematics}
\label{sec:methods:stellar_kinematics}
We extract 2D stellar kinematics using the Python package {\sc{ppxf}} \citep{ppxf_1,ppxf_2}, fitting for 4 kinematic moments; the line of sight velocity, velocity dispersion, and $h_3$ and $h_4$, the Gauss-Hermite moments which describe the skewness and kurtosis of the profile, respectively \citep{gerhard_1993,van_der_marel_1993}. To prepare the minicubes for spectral fitting, we remove all spaxels with a signal-to-noise of less than 1.5 per pixel in the fitted spectral region before Voronoi binning to a signal-to-noise of 10 per bin with the Python package {\sc{vorbin}} \citep{cappellari_adaptive_2003}. The exact fitted region changes on a galaxy-by-galaxy basis due to the variation of the sample around the $z=0.3$ target, but is approximately rest-frame $g$-band.   

We use the Indo-US template library \citep{valdes_2005_indous} due to its high spectral resolution when attempting to measure accurate velocity dispersions with the moderate redshift MAGPI data. The template library was convolved to match the wavelength-dependent line-spread function of the MUSE/MAGPI data, assumed to be spatially invariant in each field. We select an optimal subset of Indo-US templates with an initial aperture spectrum fit, for efficiency.  We restrict the individual Voronoi bin fits to the top 20 weighted templates from the aperture fit. The aperture spectrum is constructed by co-adding spaxels within the half-light ellipse from the MGE model. For all fits we use fifth degree multiplicative  polynomials.

\begin{figure}
\centering
\includegraphics[width=\linewidth]{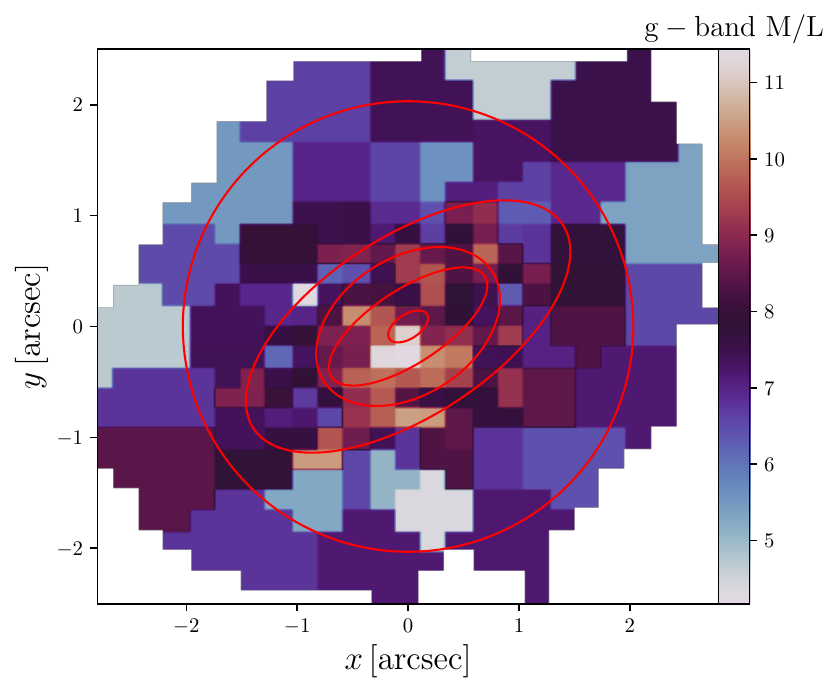}
\caption{An example stellar population Salpeter g-band mass-to-light ratio map for MAGPI object 1525170222. The overlaid red contours show the MGE components that sit within the map. The mean mass-to-light ratio along each ellipse, of width 1 pixel, is used to multiply the galaxy surface brightness model, resulting in a stellar mass model.}
\label{fig:ml_plot_example}
\end{figure}

We estimate uncertainties on the fitted kinematic moments by shuffling the {\sc{ppxf}} residuals and re-distributing them on the input spectrum before re-fitting across 100 iterations and taking the standard deviation. Since this process adds noise to a data spectrum that already contains noise, we reduce the derived parameter uncertainties by $\sqrt{2}$ to account for this noise doubling. We note this process is the same as the kinematics presented in \citet{derkenne_2023_MAGPI}, and for the same objects, except here we fit for 4 kinematic moments instead of 2, and the template broadening here is wavelength dependent. 

As the MGEs and stellar kinematics are derived from the same underlying MUSE data, we rotate the kinematic $x$ and $y$ coordinates by the photometric position angle determined from the MGE fits. This rigorously ensures that the gravitational potential described in Section~\ref{sec:methods:schwarzschild:pot} and kinematics are correctly aligned within the Schwarzschild models.

\subsection{Stellar mass-to-light ratios}
\label{sec:methods:stellar_pops}
Using the same signal-to-noise cuts and Voronoi bins as in Section~\ref{sec:methods:stellar_kinematics}, we extract stellar population measurements using the E-MILES stellar populations models of \citet{vazdekis_2016_EMILES}. These templates allow for the calculation of a spectral mass-to-light ratio whilst also accounting for variations of stellar age and metallicity. In practise, we use the included {\sc{ppxf}} E-MILES  utilities for the fit, following \citet{cappellari_2013_XX} equation 2. The mass-to-light ratios are calculated with a Salpeter \citep{salpeter_1955} IMF, however we note we leave the global scaling of mass in our Schwarzschild models free, so that resulting models are not dependent on any particular IMF. This process assumes the IMF is spatially constant within a given galaxy. 

To convert the luminous MGE model into a stellar mass model, an ellipse is constructed for each individual MGE component, using its respective $\sigma$ and $q$ values. This ellipse represents the region where that MGE component's contribution is maximal. The mean $g$-band mass-to-light ratio along this ellipse is associated with that component. All ellipses that lie beyond the coverage of the mass-to-light map are set to have the same mass-to-light ratio as the last ellipse contained within the map. That is, we assume that at large radii the mass-to-light ratio becomes constant, which avoids divergent solutions, and as seen, for example, in the detailed mass-to-light profile of NGC 5120 \citep{mitzkus_2017}, and for FCC 47 \citep{thater_2023}. Each of the $N$ Gaussian surface density components in units of $L_{\odot}/\mathrm{pc}^2$ are multiplied by their corresponding mass-to-light ratio, so that the final 2D projected galaxy  surface density model is in units of $\Msun/\mathrm{pc}^2$. This method for converting surface brightness models to stellar mass follows \citet{poci_2017_syst}. We show an example stellar mass-to-light ratio map in Figure~\ref{fig:ml_plot_example}.

We do not include uncertainties on the stellar MGE or the stellar population mass-to-light ratios, as perturbing the gravitational potential for error analysis with Schwarzschild models is prohibitively expensive in terms of computational time and energy. While the stellar mass-to-light ratio maps tend to be noisy, they do exhibit structure, and our technique of averaging the mass-to-light ratio along each of the Gaussian components helps reduce the impact of noise. The example we show in Figure~\ref{fig:ml_plot_example} is neither the worst nor best from the sample.

\section{Schwarzschild orbit-based models}
\label{sec:methods:schwarzshild_models}

Schwarzschild models are a flexible but computationally expensive method by which to create exceptionally detailed dynamical models of galaxies, recovering intrinsic orbital properties from projected radial velocity and surface density information \citep{schwarzschild_1979_numerical}. We use the triaxial {\sc{dynamite}} code \citep{jethwa_2020_dynamite} to construct the Schwarzschild models presented in this work, following the theoretical outline of \citet{bosch_2008_triaxial} with orbit mirroring corrections included \citep{quenneville_2022_triaxial,thater_2022_mirroring}. The models are computationally expensive as a large library of orbits that are allowable within a given gravitational potential are integrated for a set number of orbital periods (to ensure reliable convergence), and the resulting surface brightness and kinematic moments compared to observations. Any change to the given potential therefore results in re-integrating the orbit library to determine the orbital weights that best reproduce the observations via non-negative least squares optimisation. In practise, this can require running thousands of models to characterise a parameter space, and large orbit libraries that do not artificially restrict the various orbital structures a galaxy may have. We outline our construction of the gravitational potential, orbit libraries, and parameter space searches in the following sections. 

\subsection{Constructing the gravitational potential}
\label{sec:methods:schwarzschild:pot}
The 2D axisymmetric MGE model in units of $\Msun/\mathrm{pc}^2$ can be converted to a triaxial 3D mass density in units of $\Msun/\mathrm{pc}^3$ using the following expression:

\begin{equation}
\begin{aligned}\label{eq:3D_mass_dens}
\rho(x,y,z) = \sum_{i=1}^{N} \frac{M_i}{(\sigma_i\sqrt{2\pi})^3q_ip_i} \times \exp \left[ - \frac{1}{2\sigma_i^2}(x^2 + \frac{y^2}{p_i^2} + \frac{z^2}{q_i^2})\right],
\end{aligned}
\end{equation}
where $p_i$ = $B_i/A_i$, $q_i$ = $C_i/A_i$, and $A_i$, $B_i$, and $C_i$ are the major, intermediate, and minor axes of the 3D triaxial Gaussian. The corresponding projected quantities ($p'$, $q'$) can be transformed into the intrinsic quantities using the viewing angles of the object \citep{MGE,bosch_2008_triaxial}. $M_i$ represents the mass of the \textit{i}\textsuperscript{th} Gaussian component with dispersion $\sigma_i$.

Following \citet{zhu_2018_orbital}, for the dark matter halo we adopt a spherical Navarro-Frenk-White (NFW)  halo \citep{navarro_1997_nfw}, where the enclosed mass profile is given as 
\begin{equation}
\begin{aligned}
\label{eq:nfw}
M(<r) = M_{200}g(c)\left[\ln(1+cr/r_{200}) - \frac{cr/r_{200}}{1 + cr/r_{200}}\right],
\end{aligned}
\end{equation}
with $r$ the radius, $c$ the concentration of the halo, $g(c) = [\ln(1+c)-c/(1+c)]^{-1}$, and $M_{200}$ is the virial mass within a radius of $r_{200}$. The dark matter halo is then fully described by two free parameters: the concentration and the virial mass. However, we found, as did \citet{zhu_2018_orbital} with local Universe data, that for MAGPI data quality the two dark matter halo parameters were unconstrained in our models.  Instead we adopted an $M_{200}-c$ coupled halo.  This fixes the concentration according to the relation measured by \citet{dutton_2014_cold},

\begin{equation}
\begin{aligned}\label{eq:c_coupled}
\log_{10}c = a - b\log_{10}(M_{200}/[10^{12}h^{-1}\Msun]),
\end{aligned}
\end{equation}
with $h = 0.671$. The parameters $a$ and $b$ evolve with redshift as
\begin{equation}
\begin{aligned}\label{eq:c_coupled_a}
a = 0.520 + (0.905 - 0.520)\exp(-0.671z^{1.21})
\end{aligned}
\end{equation}
and
\begin{equation}
\begin{aligned}\label{eq:c_coupled_b}
b = -0.101 + 0.026z,
\end{aligned}
\end{equation}
given as equations 10 and 11 in \citet{dutton_2014_cold}. We fix the $M_{200}-c$ relation at the mean redshift of the MAGPI sample, {$z = 0.304$}. The critical density at this redshift with the adopted cosmology is $ \rho_c = 1.73 \times 10 ^{-7}\,\Msun \mathrm{pc}^{-3}$. Our models therefore have only one free parameter to describe the dark matter halo, $M_{200}/M_{\star}$, where $M_{\star}$ is the total stellar mass within the virial radius. For completeness, although well below the resolution of  MAGPI data, we include a black hole with black hole mass given by  the redshift dependent mass-velocity dispersion relation from \citet{robertson_2006_black}. The total mass (luminous and dark) of the  system can then be scaled by the global constant $\Upsilon_r$, as the scaled model can be simply evaluated by scaling the orbital velocities without re-integrating the orbit library.

\subsection{Building the orbit library}
\label{sec:methods:orbits}
Only certain regular orbital families can exist within any given gravitational potential. Each unique model therefore has its own library of orbits which must be numerically integrated for a sufficient number of periods so as to be representative of the time-averaged orbital properties. The orbits themselves are characterised by the three conserved integrals of motion: binding energy $E$, $z$-component angular momentum $L_z$, and the third, non-classical integral of motion, $I_3$ \citep{van_der_marel_1998,cretton_1999}.

We construct the orbit library by spanning 21 intervals in $E$, $15$ intervals in $L_z$, and $11$ intervals in $I_3$.  We choose this relatively large orbit library size based on the exploration of orbit library size conducted by \citet{quenneville_2022_triaxial}. They found that with $L_z$ sampled below 15 intervals there were spurious oscillations in the $\chi^2$ space of the model, yielding unstable solutions. However, increasing the orbit library to larger sizes does not continuously improve the models, as successively fewer orbits are given significant weighting \citep{Jin_2019_evaluating}. Our orbit library size is a compromise between large enough not  to restrict possible orbital structures, and small enough to be computationally efficient. The {\sc{dynamite}} code constructs three sets of orbit libraries: box orbits, tube-orbits, and counter-rotating tube-orbits, such that the total orbit library size is $3 \times 21 \times 15 \times 11$. We do not regularise the orbits by use of dithering, as dithering vastly increases the computational time required for each model. Dithering seeds $N^3$ orbits in a mini-grid around each set of initial conditions, where $N$ is the chosen dithering value. We explore the impact of dithering in Appendix~\ref{sec:appendix:dithering}, and conclude it is not necessary for MAGPI-like data quality. 

The minimum and maximum orbital radius is set as $0.5\sigma_{\mathrm{min}}$ and $5\sigma_{\mathrm{max}}$, where $\sigma_{\mathrm{min}}$ and $\sigma_{\mathrm{max}}$ are the minimum and maximum Gaussian dispersions from the MGE fit, respectively, following \citet{zhu_2018_orbital}. These radii set the minimum and maximum orbital energy within a given gravitational potential. We integrate orbits across 200 orbital periods, with 50000 samples per orbit in the meriodional plane. Again following \citet{zhu_2018_orbital}, we implement the `CRcut' keyword in {\sc{dynamite}}, which avoids the spurious up-weighting of counter-rotating orbits for galaxy regions with high $v/\sigma$.

The observable properties of each orbit, stored on an intrinsic coordinate grid, are transformed to projected coordinates. The orbital properties (surface brightness and higher order stellar kinematic moments) are then randomly perturbed by the PSF before being assigned to the observational apertures (the Voronoi bins). In this way, the intrinsic models are convolved with the same PSF as the observational data \citep{bosch_2008_triaxial}. 

\begin{figure}
\centering
\includegraphics[width=\linewidth]{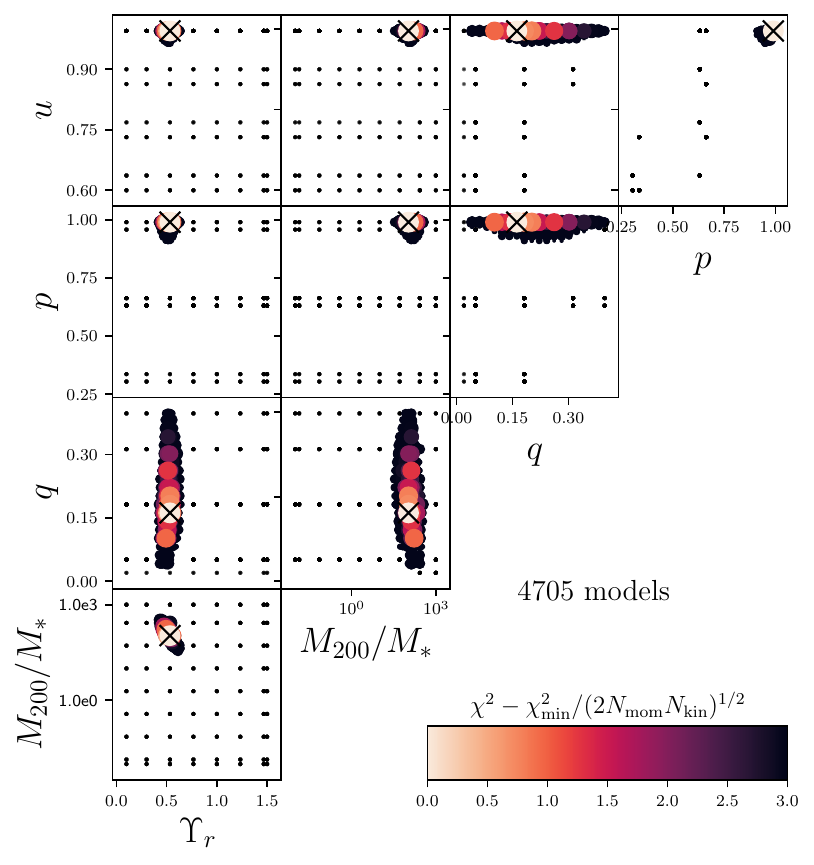}
\caption{A corner plot of the free parameters in the Schwarzschild models for MAGPI object 1525170222; the intrinsic axial ratios $p$, $q$, and $u$, the dark matter fraction at $r_{200}$ labelled as $M_{200}/M_{\star}$, and the global mass re-scaling, $\Upsilon_r$. Small black dots indicate points outside the 3-sigma region. Larger points indicate models within the 3-sigma region, coloured by their offset from the best-fit model as judged by the chi-square value, calculated as $\chi^2 - \chi^2_{\mathrm{min}}/(2N_{\mathrm{mom}}N_{\mathrm{kin}})^{\frac{1}{2}}$, with $N_{\mathrm{kin}}$ the number of Voronoi bins and $N_{\mathrm{mom}}$ the number of kinematic Gauss-Hermite moments. The best-fit model is shown with a black cross. All 4075 models are shown, which generally approximates the number of models run for each MAGPI object. This example is representative of the constraint on the parameters for the models.}
\label{fig:param_space}
\end{figure}

\subsection{Parameter space search and model selection}
\label{sec:methods:parameter_space}

To summarise the above, our implementation of the Schwarzschild models include five free parameters:

\begin{enumerate}
    \item The axial ratio $p = B/A$, where $B$  and $A$ represent the medium and major axes of the 3D triaxial Gaussian, respectively.
    \item The axial ratio $q = C/A$, where $C$ is the minor axis of the 3D triaxial Gaussian.
    \item The axial ratio $u = A'/A$, where $A'$ is the projected major axis of the 3D triaxial Gaussian.
    \item The global mass scaling factor, which scales both stellar and dark components jointly, $\Upsilon_r$.
    \item The dark matter content, parameterised as $M_{200}/M_{\star}$, where $M_{200}$ is the total mass within the virial radius $r_{200}$, and $M_{\star}$ is the total stellar mass.
\end{enumerate}

The values of $p$, $q$, and $u$ are all bounded such that a valid deprojection of the 2D stellar mass model exists. These conditions are that $0 < q \leq p \leq 1$; $\mathrm{max}(q/q_{\mathrm{obs}}, p) < u$ with $q_{\mathrm{obs}}$ the minimum axial ratio from the stellar surface brightness MGE model; and  $u <\mathrm{min}(p/q_{\mathrm{obs}}, 1)$. Due to these conditions, the exact bounds used in the models vary for each galaxy.

We first run what we call a `coarse' grid of models across the allowed parameter space for the five free parameters, sampling a minimum of 3 evenly spaced values within the bounds for $p$, $q$, and $u$, at least 6 values in $\Upsilon_r$ between 0.1 and 1.5, and at least 7 values in dark matter fraction $M_{200}/M_{\star}$ between -2 and 3 in log space.   

\begin{figure*}
\centering
\includegraphics[width=0.8\linewidth]{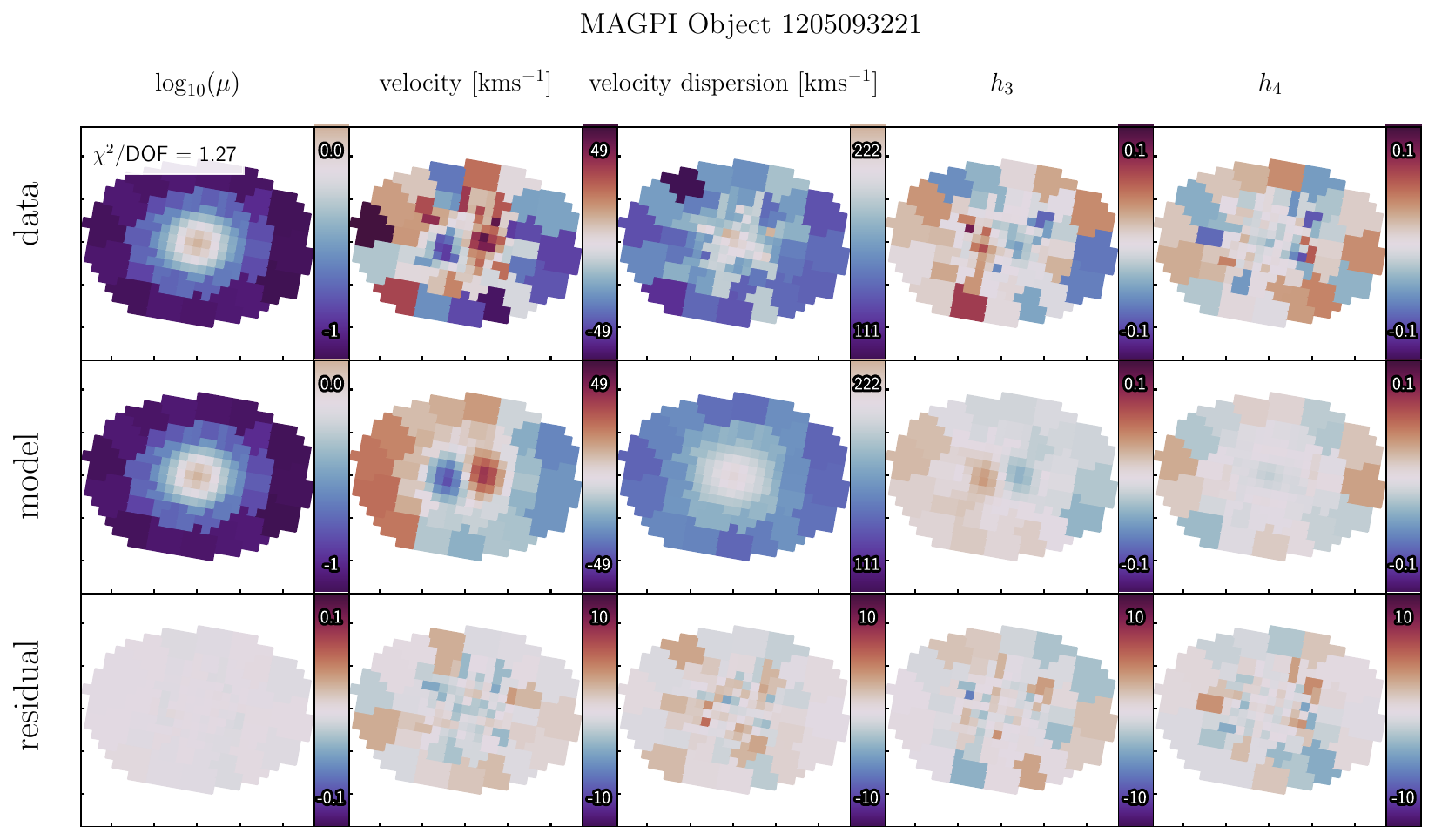}
\caption{Top row: The observed data, with the reduced $\chi^2$ value inset in the top left corner of the surface brightness panel, $\log_{10}(\mu)$. Note the distinct core feature in the velocity field. Middle row: The best-fit Schwarzschild model, judged by the $\chi^2$ value. Bottom row: The re-scaled residuals, given as $(\mathrm{data}-\mathrm{model})/\Delta \mathrm{data}$, where $\Delta \mathrm{data}$ are the data uncertainties. The fields are shown rotated by the kinematic position angle of the velocity map. The x- and y-axis ticks are 1 arcsecond intervals.}
\label{fig:1205093221_kinematics}
\end{figure*}

Each resulting model is compared to the kinematic observations to determine the best-fit model via the chi-square value, calculated from the observed kinematics and the Schwarzschild model projected kinematics. The  galaxy surface brightness models are used to constrain the linear combination of orbits, but are not used to evaluate the quality of fit between different models, which is done using only the kinematic moments. The best-fit model from the coarse run is then used to initiate a dynamic `fine' parameter search with increasingly smaller step sizes around the best-fit solution, up to a minimum of $0.05$ in log space for dark matter fraction, 0.02 for $p$ and $q$, 0.01 for $u$  (which generally tends very close to unity), and 0.01 in $\Upsilon_r$. Practically, the coarse and fine model runs use the `FullGrid' and `LegacyGridSearch' options in the {\sc{dynamite}} code, respectively. We allow the iterations to continue for up to 6000 coarse models and 2000 fine models before terminating, although the code can terminate sooner once the $\chi^2$ of new models makes no further improvement, quantified as an absolute reduction in the $\chi^2$ value of less than 0.1.

As in \citet{zhu_2018_orbital}, we define the $1\sigma$ confidence level as $\Delta \chi^2 = \sqrt{2N_{\mathrm{kin}}N_{\mathrm{mom}}}$, where $N_{\mathrm{kin}}$ is the number of kinematic Voronoi bins and $N_{\mathrm{mom}}$ is the number of kinematic Gauss-Hermite moments. The coarse and fine modelling runs are combined to establish all models within the $1\sigma$ confidence region. The quoted parameter value (e.g., $p$) is calculated from the model with the lowest $\chi^2$ value. To estimate the uncertainties, parameter values from all $1\sigma$ models are calculated, and the upper and lower bounds of that distribution taken. For example, the upper error is quoted as $\mathrm{max}(\textbf{X}) - x_0$, where $\textbf{X}$ is the vector of all $1\sigma$ model parameter values and $x_0$ is the value from the lowest $\chi^2$ model.  An example of the parameter space coloured by Schwarzschild model $\chi^2$ value is shown in Figure~\ref{fig:param_space}. 

\begin{figure*}
\centering
\includegraphics[width=0.8\linewidth]{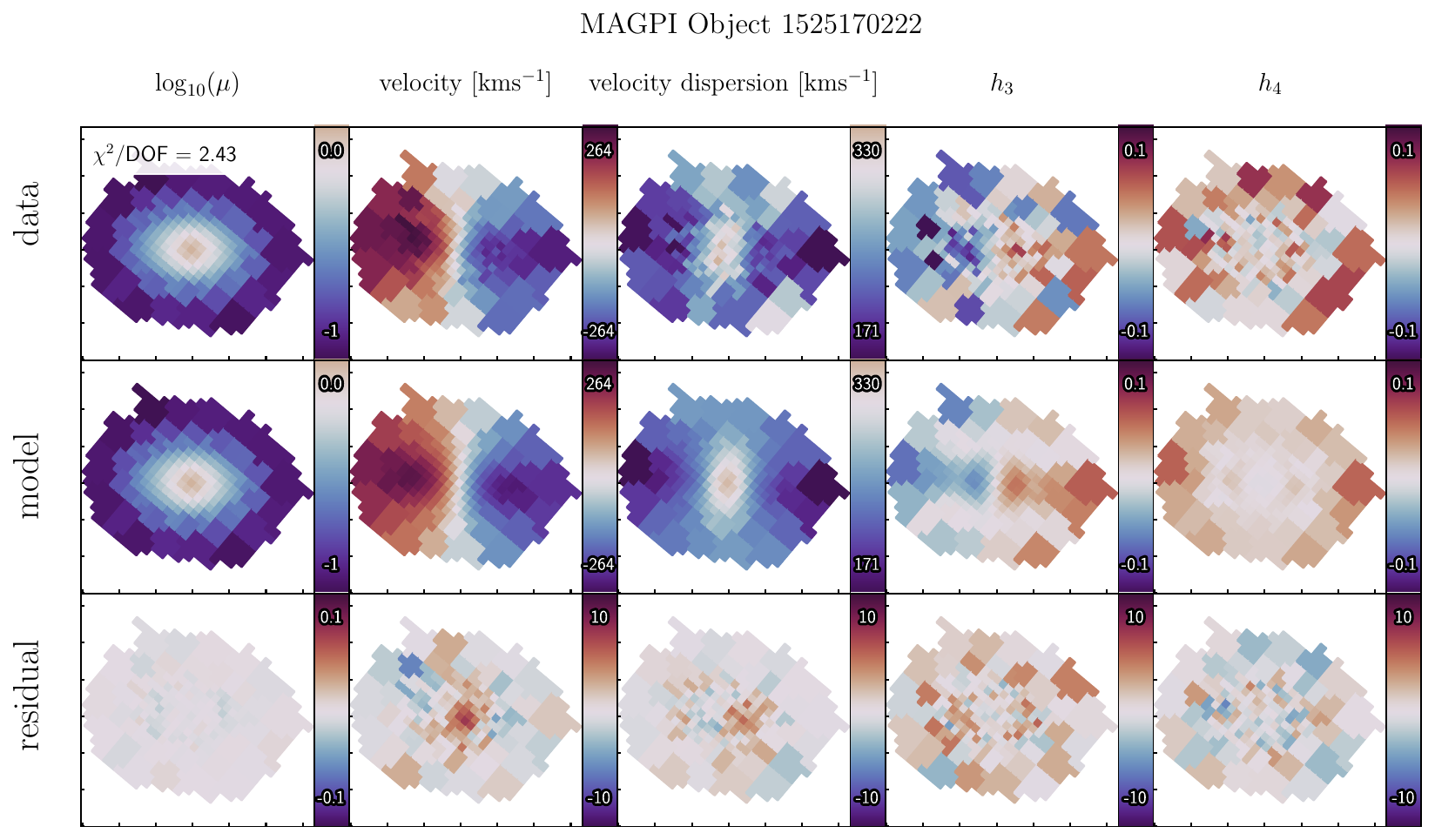}
\caption{The same as Figure \ref{fig:1205093221_kinematics}, except for MAGPI object 152517022. We display this example in particular because it has one of the highest best-fit model reduced $\chi^2$ value of the sample, to show the general quality of models in comparison to the observed data.}
\label{fig:1525170222_kinematics}
\end{figure*}

Finally, we assess the quality of the resulting models. As noted above, the model iterations were informed by the kinematic $\chi^2$ value only, and so there is no requirement for the best-fit model's surface brightness distribution to match the observations. In some cases, the kinematics were well reproduced but not the surface brightness, which in turn means the galaxy mass distribution is not well represented by the model. We therefore select only models where the median absolute deviation of the surface brightness residuals (data-model/data) was less than 1 per cent, ensuring only galaxies with near structureless surface brightness residuals are included in the final sample. This strict cut leaves 22 out of a possible 30 objects. We note the cut galaxies are generally spiral systems with bars or kinematic twists, leaving a sample that is dominated by early-type systems. We show two example Schwarzschild models in Figures~\ref{fig:1205093221_kinematics} and  \ref{fig:1525170222_kinematics}, which are representative of the selected sample. We show MAGPI object 1525170222 in particular because it has one of the highest reduced $\chi^2$ values of the sample at 2.43 (i.e., one of the worst fits) but a model that well represents the observed kinematics, to demonstrate the general fit quality of the models. The remaining 20 Schwarzschild models are all shown in Appendix~\ref{sec:appendix:models}.

\section{Results}
\label{sec:results}
From our Schwarzschild models we can extract detailed, intrinsic orbital structures of galaxies. In the following sections we present our analysis of the \Nsample MAGPI galaxies with well-fitting Schwarzschild models. We first examine the bulge-halo conspiracy with our data, showing in particular the large scatter of total density profile slopes and how the dark matter fractions are independent from the stellar mass structure. We then present an analysis on the intrinsic shapes and orbital structures of the sample. Finally we present a comparison of our total mass density profile slopes from the Jeans models presented in \citet{derkenne_2023_MAGPI} and Schwarzschild models derived in this work.

\subsection{Testing the bulge-halo conspiracy}
\label{sec:results:bulge_halo}

We examine two hypotheses of the bulge-halo conspiracy. The first is that populations of galaxies have low intrinsic scatter in their total mass density profile structure, quantified through the total mass density profile slope. From this, we would expect the observed scatter in total mass density slopes to be less than or equivalent to the observed scatter in the stellar mass density profile slopes. Secondly, if the dark and baryonic components conspire through some as of yet unknown mechanism, then the dark matter content of galaxies should correlate with the stellar density structure.  

\begin{figure*}
\centering 
\includegraphics[width=0.9\linewidth]{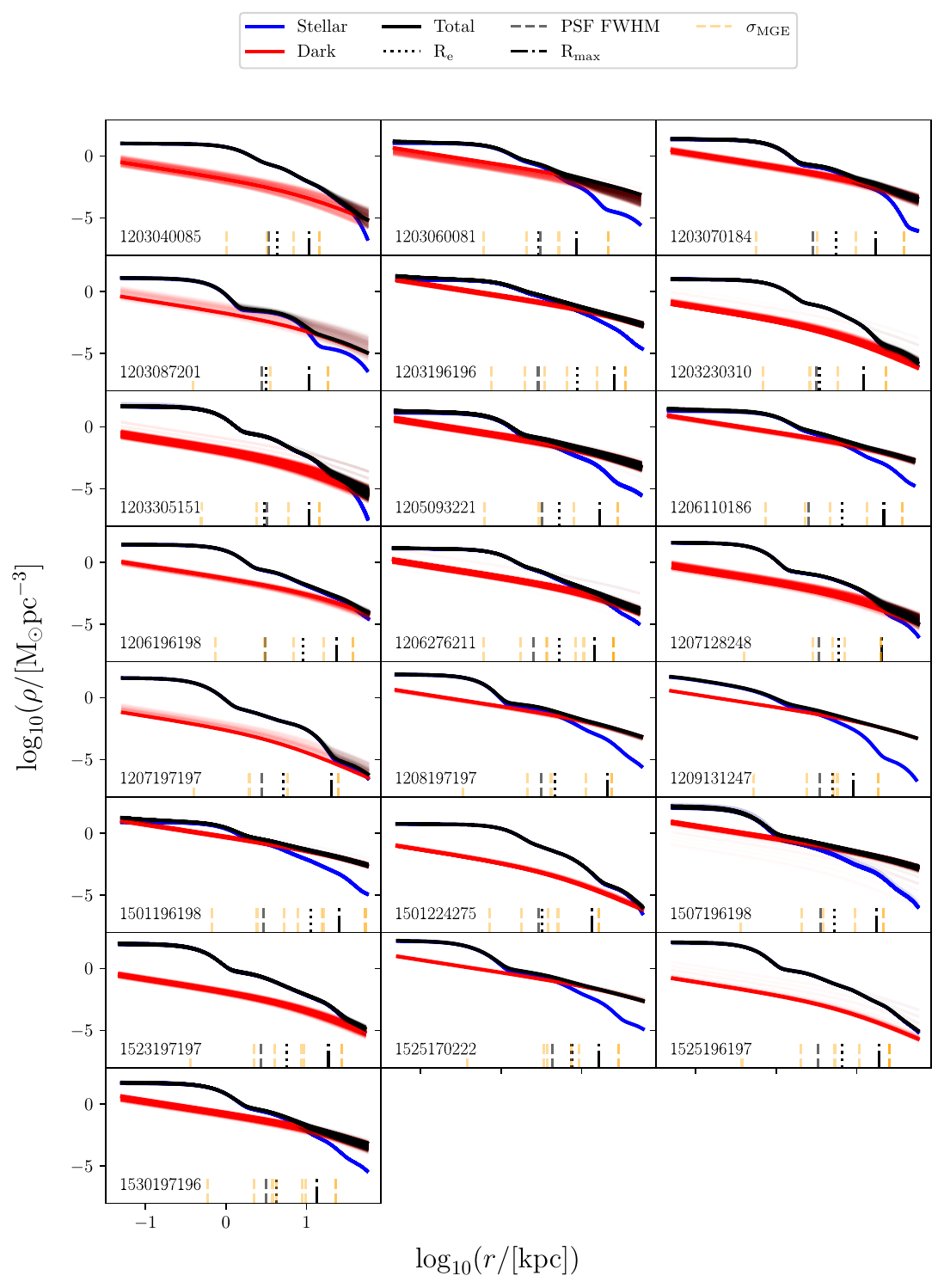}
\caption{All the total (black), stellar (blue) and dark matter (red) density profiles for the sample. All the profiles from the $1\sigma$ Schwarzschild models are shown. The total profiles mostly obscure the stellar profiles, except for at large radii. The dashed black line shows the PSF FWHM, the dotted line shows the half-light radius, and the dot-dashed line shows the maximum extent of the kinematic data. These profiles are intrinsic and do not include the effects of a PSF. The orange dashed lines show the Gaussian sigmas of the MGE surface density models. The MAGPI ID of each object is inset in the lower left corner of each panel.}
\label{fig:all_profiles}
\end{figure*}

The structure and gradient of the stellar mass profile is derived using the stellar mass models described in Section~\ref{sec:methods:stellar_pops} (i.e. the mass-to-light adjusted MGE models), scaled by the best-fit parameter $\Upsilon_r$ from the Schwarzschild models. The dark matter profile is determined by the fitted dark halo in the Schwarzschild models, with the total profile being the simple addition of the stellar and dark components. The total, stellar, and dark matter mass density profiles for the whole sample are shown in Figure ~\ref{fig:all_profiles}. 

Following the approach of previous works \citep[e.g.][]{cappellari_2015_small}, we measure the stellar and total mass density slopes - referred to as $\gamma_{\star}$ and $\gamma_{\mathrm{tot}}$, respectively - assuming a single power law of the form:
\begin{equation}
    \gamma = \frac{d\log(\rho)}{d\log(r)},
\label{eq:define_gamma}
\end{equation}
where $\rho$ is the mass density and $r$ is the galactocentric radius. As in \citet{derkenne_2023_MAGPI}, we fit the slope for the radial region between \re/10 and 2\re.

\begin{figure}
\centering
\includegraphics[width=0.95\linewidth]{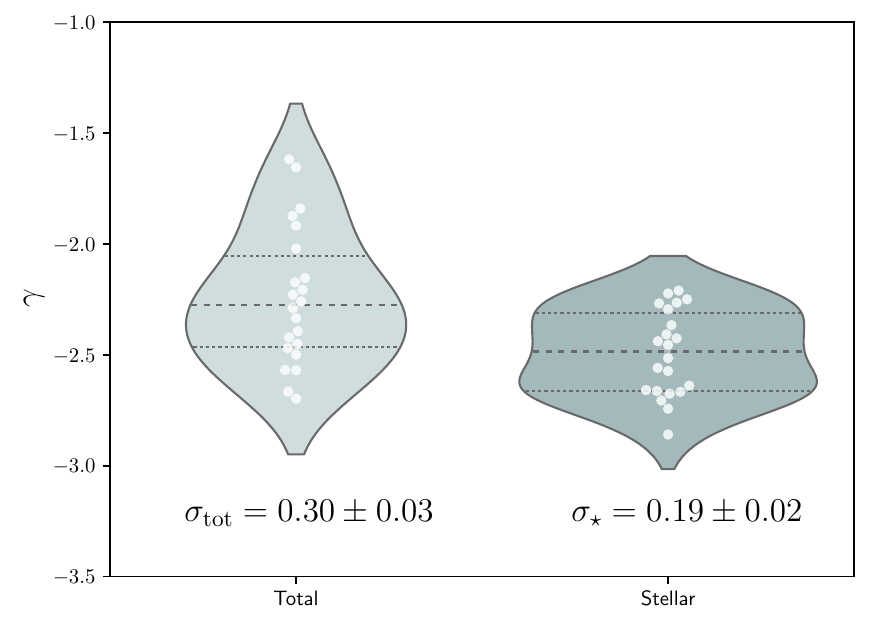}
\caption{Violin plots with inset data swarm (white dots) and Gaussian kernel density  estimate (coloured background) of the total density slopes compared to the stellar density slopes of the sample. The dashed  lines indicate the sample medians, and the dotted lines indicate the sample quartiles. The standard deviation for each slope is inset. The total slopes have a higher standard deviation than the stellar slopes, which is not consistent with the bulge-halo conspiracy.}
\label{fig:swarm}
\end{figure}

From the power law fits to the MAGPI sample mass density profiles we examine the total profile and stellar profile slope distributions, with the results shown in Figure~\ref{fig:swarm}. The stellar slopes are steeper on average than the total profile slopes, as the baryonic component dominates the central regions but not the outskirts, and the inclusion of a dark matter halo tends to cause the combined profile to become more shallow. We note that the scatter of the total density profiles is  larger than for the stellar component alone, comparing $\sigma_{\mathrm{tot}} = 0.30 \pm 0.03$ to $\sigma_{\star} = 0.19 \pm 0.02$ (errors here are the standard error on the standard deviation). We find a super-isothermal median total mass density profile slope of $\gamma_{\mathrm{tot}} = - 2.27 \pm 0.08$ (with the uncertainty characterised as the standard error on the median), and the sample spans $\gamma_{\mathrm{tot}} \sim - 1.6$ to $\gamma_{\mathrm{tot}} \sim -2.7$ in slope values. For the stellar component, we find a median slope of $\gamma_{\star} = -2.28 \pm 0.05$ (standard error of the median) and the sample spans between $\gamma_{\star} \sim -2.2$ to $\gamma_{\star} -2.85$. As might naively be expected, the spread of values in the total mass density profile slopes, which combines the stellar light structure with a free dark matter halo, is higher than the spread of the stellar mass profiles alone. By contrast, the bulge-halo conspiracy predicts reduced or equivalent scatter when considering the total mass density profile slopes against the stellar mass profile slopes.

The variability of the total mass density profile slopes presented in this work is in contrast to some previous works. Gravitational lensing studies generally report a scatter of $\lesssim 0.2$, such as \citet{barnabe_two_2011},  \citet{Li_2018_strong_lensing}, and \citet{etherington_2023_beyond}. However, \citet{ruff_sl2s_2011} find comparable scatter to our results with $\sigma_{\gamma} = 0.25$. Several dynamics studies have also found comparatively low scatter, with \citet{cappellari_2015_small} finding $\sigma_{\gamma} = 0.11$ and \citet{bellstedt_sluggs_2018} finding $\sigma_{\gamma} = 0.13$, but \citet{li_manga_2019} finding a  higher value of $\sigma_{\gamma} = 0.22$. \citet{poci_2017_syst} explicitly compared the slope distributions between the stellar mass profile slopes and total mass density slopes and found equivalent scatter for the \atlas sample, and could not refute the bulge-conspiracy on that basis. However, the \atlas data typically extends to only a half-light radius, which limits how well the dark matter halo can be constrained. \citet{poci_2017_syst} found $\sigma_{\gamma} \sim 0.17$ for both the stellar and total mass density profile slope distributions. 

\begin{figure}
\centering
\includegraphics[width=\linewidth]{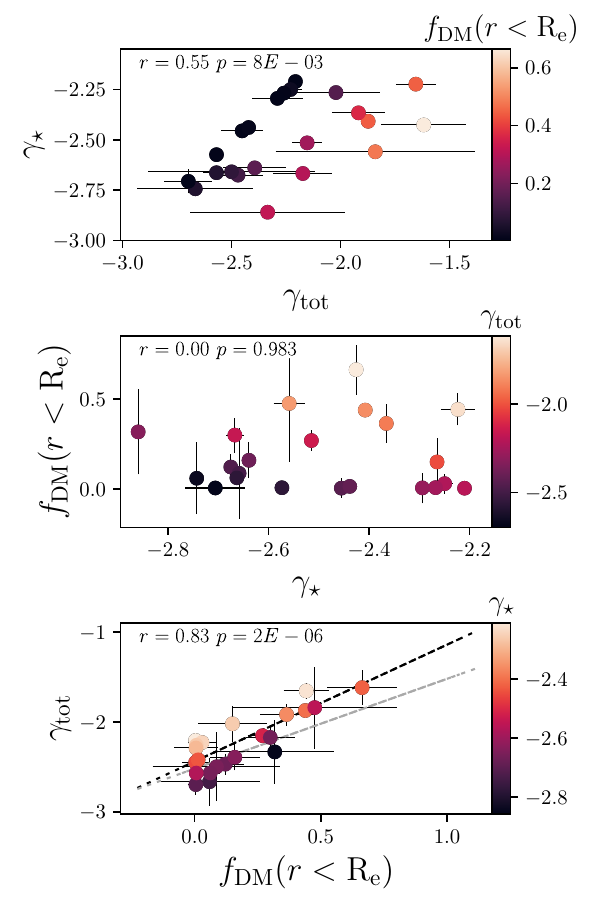}
\caption{Scatter plots of all variable pairs of the total slope, $\gamma_{\mathrm{tot}}$, the stellar slope, $\gamma_{\star}$, and the dark matter fraction within the half-light radius, $f_{\mathrm{DM}}(r < \re)$. Each plot is coloured by the third variable. The Pearson r and p-values are inset. In the bottom panel, a  fit to the MAGPI galaxies is shown as a black dashed line, and the relation for Magneticum galaxies shown as a grey dashed line, with the relation taken from \citet{rhea_2017_coevolution}.} 
\label{fig:total_stellar_DM}
\end{figure}

We can also test the correlations between the total profile slopes, the dark matter content within the half light radius, and the stellar profile slopes. We show these results in Figure~\ref{fig:total_stellar_DM}. We note our results are unchanged if we consider the dark matter profile slopes instead of dark matter fractions. We find a moderate correlation between the stellar mass density profile slope and the total mass density profile slope, in that a galaxy  with a steep stellar component (where the stellar mass density drops off quickly with radius) also has a steep total component, with a Pearson correlation coefficient of 0.55, implemented through {\sc{scipy}} \citep{2020SciPy-NMeth}. This makes sense, as the inner regions of the total slope are generally just the stellar component. For objects with no dark matter, the total mass density slope must equal the stellar mass density slope by construction. The total density profile slopes are strongly correlated with the dark matter fractions within the half-light radius, with a correlation coefficient of 0.83, such that shallower total profiles have a higher dark matter fraction.  However, we notice the lack of even a mild correlation between the half-light dark matter fraction and the stellar profile slope, with a correlation coefficient of 0 and p-value of 0.98. This result indicates that the stellar mass structure of a galaxy is independent of its dark matter content, again at odds with the supposition of a `conspiracy' between the stellar and dark matter components. 

From our results, we argue that the accumulation of the dark matter halo perturbs the total density slope from steep to shallow values, but that the stellar mass structure does not determine the structure of the dark halo (or the reverse).  That the total slope correlates with dark matter fraction is a prediction from the Magneticum \citep{dolag_2015_magneticum} simulations. \citet{rhea_2017_coevolution} found that the total slope and dark matter  fraction co-evolves for all systems across all redshifts, with a simple relation of the form $\gamma_{\mathrm{tot}} = 1f_{\mathrm{DM}} -2.52$ with $f_{\mathrm{DM}}$ the dark matter fraction within the 3D half-mass radius. With the MAGPI data, we find a notably similar result, of $\gamma_{\mathrm{tot}} = (1.3 \pm 0.2) f_{\mathrm{DM}} - (2.44 \pm 0.04)$, with $f_{\mathrm{DM}}$ the dark matter fraction within the half-light radius. Intuitively, the relation between the two parameters makes sense. Galaxies with low dark matter content have total slopes that closely follow the stellar mass distribution. Subsequent (dry) mergers increase the dark matter fractions by increasing the effective radius of the galaxy with stars at large radii, causing the total slope to approach shallower (less negative) values as the extent of the galaxy encompasses more of the (shallower) dark matter profile.

The lack of correlation between the dark matter content and stellar mass structure is in tension with some previous observations. \citet{tortora_2018_last} investigated the relationship between central dark matter fractions and stellar density using aperture-based dynamical masses from Sloan Digital Sky Survey (\citet{abazajian_2009_SDSS}, SDSS) spectroscopy, and structural parameters from Kilo Degree Survey (\citet{de_jong_2013_kilo}, KiDS) photometry for a sample of $\sim 3800$ galaxies spanning up to $z \sim 0.65$, and found  that there exists a strong anti-correlation between the dark matter content of a galaxy and its average central density. Galaxies with the steepest S\'{e}rsic indices, of $n\sim10$, were found to have the largest central dark matter fractions. Likewise, an earlier study by \citet{grillo_2010} with a sample of $1.7\times10^{5}$ SDSS galaxies found a mild anti-correlation between central dark matter fractions and stellar surface densities, assuming an isothermal power-law density profile.

While it is uncertain what exactly causes the disagreement between those studies and this work, we note that in this work we examine stellar \textit{mass} profiles instead of stellar light structure, and the Schwarzschild modelling technique implemented applied here to IFS data makes fewer assumptions than aperture based measurements of galaxy parameters. We observe no significant correlation between $f_{\mathrm{DM}}$ and S\'{e}rsric indices with the MAGPI sample, using single-component indices calculated using {\sc{galfit}} \citep{Peng_2002_galfit}. We obtain a p-value of 0.34 and a correlation coefficient of 0.34, in keeping with our results between the lack of conspiracy between the stellar component and central dark matter fractions.

\subsection{The dark matter content of massive galaxies}
\label{sec:results:dark_matter}

\begin{figure}
\centering
\includegraphics[width=\linewidth]{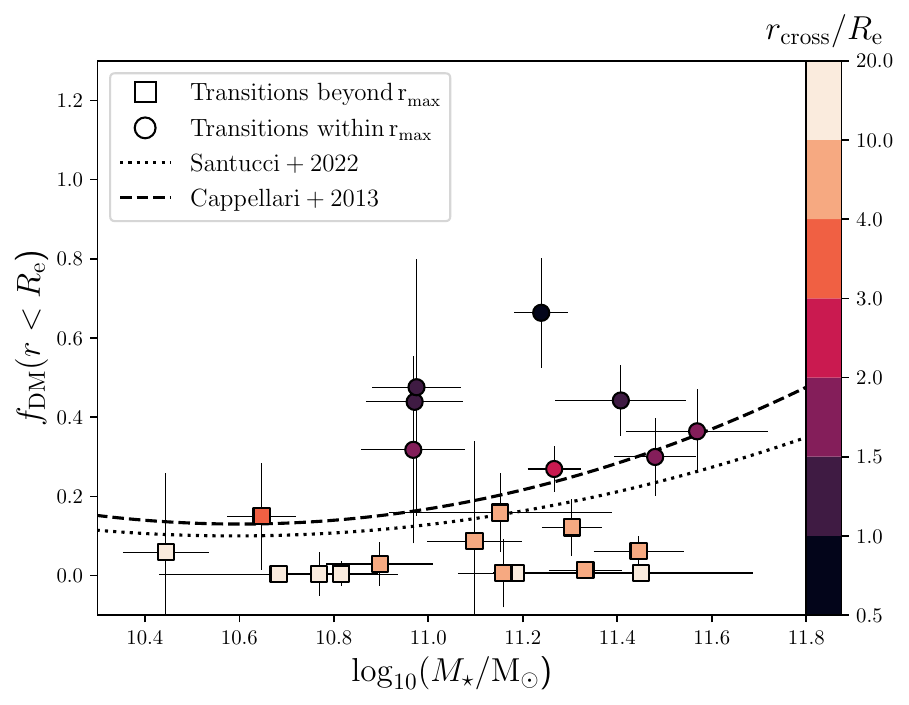}
\caption{The dark matter fractions within the half-light radius, $f_{\mathrm{DM}}(r<\re)$, as a function of stellar mass. The points are coloured by the radius at which the dark halo density exceeds the stellar mass density. The fitted relationships between dark matter fractions and stellar mass are also shown, from \citet{cappellari_2013_benchmark} and \citet{Santucci_2022}.}
\label{fig:DM_fractions}
\end{figure}

The Schwarzschild models of the MAGPI sample allow us to investigate in detail where each mass component, baryonic or dark, dominates. In Figure~\ref{fig:DM_fractions} we show the half-light dark matter fractions as a function of stellar mass, and coloured by the radius, $r_{\mathrm{cross}}$, where the dark matter density profile exceeds the stellar profile.  As with the total density profiles themselves, we find that there is a spread in values for where dark matter becomes the dominant density component. The smallest crossing radius in the sample is MAGPI object 1501196198, where the dark matter density exceeds the stellar density within the half-light radius, and has a dark matter fraction within the half-light radius of $66_{-8}^{+19}$ per cent. Numerous galaxies experience no transition within the data region and have negligible central dark matter fractions, but from the best-fitting Schwarzschild models have profiles that cross over in excess of 10-20 \re, whereas other galaxies  are dominated by dark matter in the 1-4 \re region. We conclude from this that assuming a common scale radius across a population of galaxies at which the dark matter dominates is problematic, and probably not reflective of realistic galaxy structures. This result is consistent with analysis performed by \citet{harris_2020_measuring}, who found remarkable variety in the dark matter content of a sample of 102 early-type galaxies up to 5 \re using X-ray gas to probe the density profiles. In their study, some systems are entirely dominated at 5 \re by dark matter ($f_{\mathrm{DM}} > 0.95$), and others have very little even at this large radius ($f_{\mathrm{DM}} < 0.3$).

For the MAGPI sample, the median dark matter fraction within the half-light radius is 10 per cent with a large sample standard deviation of 19 per cent. This is similar to local Universe measurements from \atlas and `Mapping Nearby Galaxies at Apache Point Observatory' \citep[MaNGA,][]{bundy_2015_manga} galaxies made using Jeans dynamical modelling, which depending on the dark matter halo adopted, sit between 8 and 17 per-cent, with lower dark matter fractions reported for galaxies with higher data quality \citep{cappellari_2013_benchmark,zhu_2024_manga}. \citet{Santucci_2022} used Schwarzschild dynamical modelling on a sample of SAMI galaxies to determine their central dark matter fractions, the closest study in methodology to this work, finding a median of 28 per cent with a standard deviation of 20 per cent. However, in our models we have accounted for radially-varying stellar population mass-to-light ratios, which might drive some of the observed differences in dark matter fractions. The fitted relations between stellar mass and dark matter fractions as found by \citet{Santucci_2022} and \citet{cappellari_2013_benchmark} are shown in Figure~\ref{fig:DM_fractions}.

We note that in our analysis we have left the global mass scaling parameter, $\Upsilon_r$, free to vary for each galaxy. This can effectively account for stellar IMF variations between galaxies (as well as any other residual mass normalisation) as reported in the literature \citep{cappellari_2012_IMF,smith_2020_evidence}; however it does not account for IMF variations \textit{within} galaxies. Since the assumed IMF can adjust the stellar mass distribution, we explore the possible impact of a radially varying IMF in Appendix~\ref{sec:appendix:imf}, which yield further uncertainties on our reported dark matter fractions of order 10 per cent. In future, and with data that can provide constraints on the IMF, a more complete treatment would include a fully fitted, 2D IMF as part of the modelling process.

\subsection{Orbital structure and intrinsic shapes of massive galaxies}
\label{sec:results:orbits_and_shapes}

\begin{figure}
\centering
\includegraphics[width=0.95\linewidth]{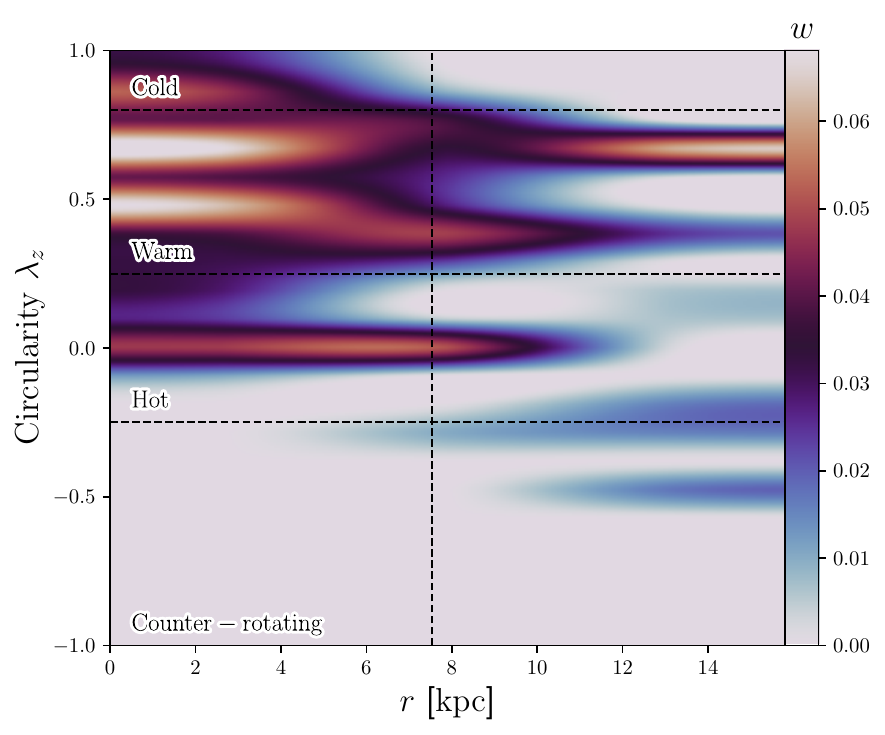}
\caption{The circularity space as a function of radius for MAGPI object 1525170222. The horizontal dashed lines show the transition between counter-rotating, hot, warm, and cold orbits. The vertical dashed line shows the half-light radius, and the $x$-axis extends to the maximum kinematic extent of the data. The image is coloured by the orbital weights, $w$, which have been normalised within  a grid of $7 x 21$ bins and interpolated for plotting purposes. MAGPI object 1525170222 has mainly warm orbits, which contribute an orbital weight fraction of $0.59^{+0.07}_{-0.05}$ within the half-light radius.}
\label{fig:circularity}
\end{figure}

The weighting of each orbit in the gravitational potential of the best-fit Schwarzschild model allows us to examine the orbital structure of galaxies, through the `circularity' space as a function of radius. The circularity, $\lambda_z$, is defined as 
\begin{equation}
\begin{aligned}\label{eq:circularity}
\lambda_z = \overline{L_z}/(r \times \overline{V_c}),
\end{aligned}
\end{equation}
with $\overline{L_z} = \overline{xv_y - yv_x}$, where $v$ represents the instantaneous velocity in a coordinate direction ($x$, $y$, $z$) and $\overline{L_z}$ is the time-average $z$-component of the angular momentum. Furthermore, $r = \overline{\sqrt{x^2 + y^2 + z^2}}$, and $\overline{V_c} = \sqrt{\overline{v_x^2 + v_y^2 + v_z^2+ 2v_xv_y + 2v_xv_z + 2v_yv_z}}$.     
 From this,  a `cold' circular orbit in an ordered disc has $\lambda_z = 1$. A `hot' box or radial orbit has $\lambda_z = 0$. Following \citet{Santucci_2022}, we define counter-rotating orbits as having $\lambda_z \leq - 0.25$, hot orbits as $-0.25 < \lambda_z \leq 0.25$, warm orbits as $0.25 < \lambda_z \leq 0.8$, and cold orbits as $\lambda_z > 0.8$. An example of the circularity space can be seen in Figure~\ref{fig:circularity} for MAGPI 1525170222, which is dominated by warm orbits. From the cuts defined above, we calculate the orbital fractions  for the sample within \re. 

\begin{figure}
\centering
\includegraphics[width=0.95\linewidth]{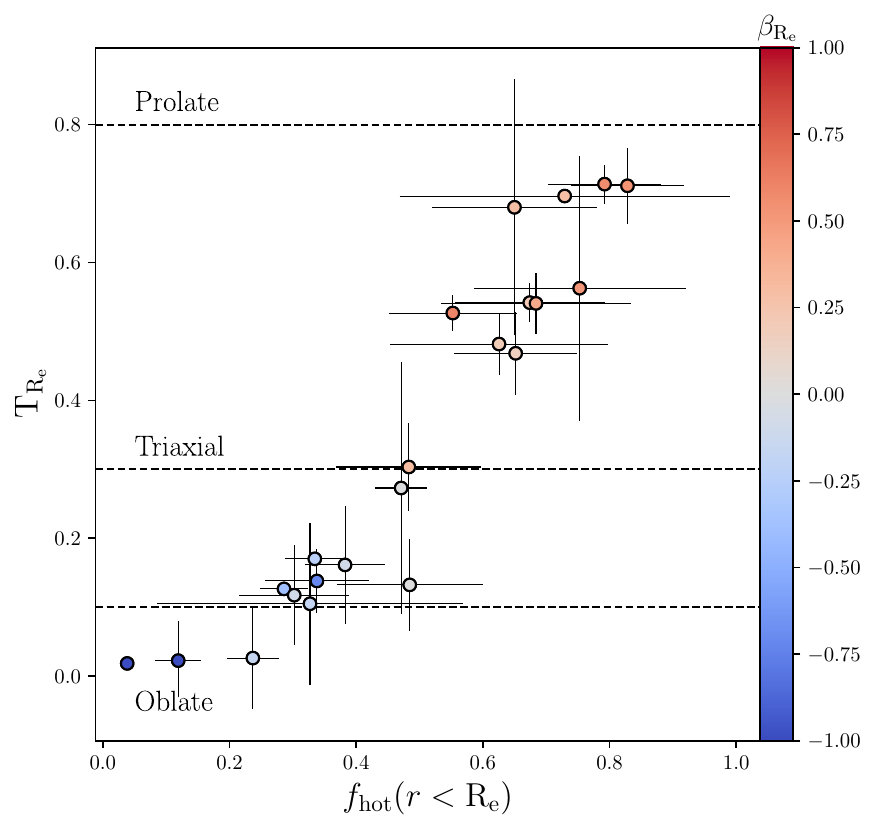}
\caption{The triaxiality, \tr, of the sample against the fraction of hot orbits within the half-light radius, where hot orbits are between -0.25 and 0.25 in circularity. The points are coloured by the anisotropy measured at the half-light radius, $\beta_{\re}$. Dashed lines indicate the boundaries between oblate, mildly triaxial, triaxial, and prolate shapes at 0.1,0.3, and 0.8 in triaxiality, respectively, following the bounds adopted by \citet{Santucci_2022}. There exists a strong correlation between radially anisotropic systems and triaxial intrinsic shapes, with a Pearson r-value of 0.74.}
\label{fig:triax}
\end{figure}

We examine the MAGPI sample for correlations between the orbital fraction (hot, warm, cold, and counter-rotating) within a half-light radius and other galaxy properties. We find a significant correlation between the intrinsic shape of a galaxy and the hot orbit fraction, shown in Figure~\ref{fig:triax}. We parameterise the intrinsic shape of a galaxy through its triaxiality parameter, defined as \tr. The shape of a galaxy is not held constant with radius in Schwarzschild models, and so we take the values of $p$ and $q$ at the half-light radius for the triaxiality calculation. We find that the triaxiality of the sample also correlates strongly with the orbital anisotropy at the half-light radius, where we define the anisotropy as 
\begin{equation}
    \begin{aligned}\label{ef:anisotropy}
    \beta_{r} = 1 - \frac{\sigma_t}{\sigma_r},
    \end{aligned}
\end{equation}
where $r$ is the intrinsic galaxy radius and $\sigma_t = \sqrt{(\sigma_{\phi}^2 + \sigma^2_{\theta})}/2$, with  $\sigma_r$, $\sigma_{\phi}$, and $\sigma_{\theta}$ the radial, azimuthal, and polar angular velocity dispersions in spherical coordinates, respectively. We choose this definition of anisotropy based on the results of \citet{thater_2022_cross}, who found that the velocity ellipsoids for NGC 6958 are  more closely aligned with spherical coordinates than the commonly adopted cylindrical coordinates. In our MAGPI sample, radially anisotropic ($\beta_r > 0$) objects are likely to be intrinsically triaxial and have mainly hot (box or radial) orbits, with few cold (short-axis tube)  orbits.

\begin{figure}
\centering
\includegraphics[width=\linewidth]{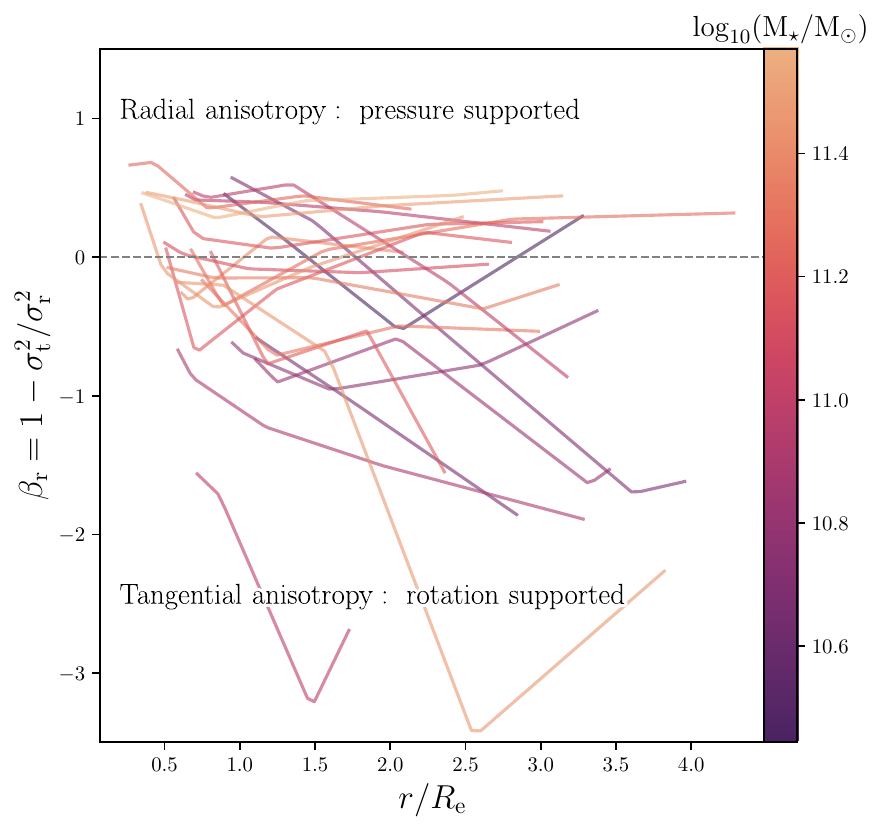}
\caption{The anisotropy of the models shown as a function of radius and coloured by stellar mass. Each line represents a MAGPI galaxy, plotted between the PSF FWHM in kpc and the maximum kinematic extent of the data. The median error on the anisotropy at 1\re for the sample is 0.11, with a maximum error of 0.24. The horizontal line at an anisotropy of zero shows the boundary between pressure and rotation supported systems. Massive systems tend to have radial anisotropy that is approximately constant with radius or mildly increasing, however many systems also exhibit decreasing anisotropy profiles. At 1 \re the typical error is $0.11$. The two objects with the steep profiles and tangential anisotropy at large radii are MAGPI objects 1207197197 and 1209131247.}
\label{fig:beta_profiles}
\end{figure}

We show the anisotropy profiles for the sample in Figure~\ref{fig:beta_profiles}, plotted between the PSF FWHM and maximum kinematic extent of the data, where we denote the anisotropy calculated at the half-light radius as $\beta_{\re}$. A common practise for constructing dynamical and lensing models of galaxies is to either hold orbital anisotropy as a global, albeit free, constant \citep{cappellari_2013_benchmark,poci_2017_syst,bellstedt_sluggs_2018,derkenne_2023_MAGPI}, or to assume the systems are isotropic \citep{auger_sloan_2010,ruff_sl2s_2011,sonn_sl2s_2013}. For the MAGPI sample, we find instead that only the massive galaxies in the sample, with $\log_{10}M_{\star}/M_{\odot} \gtrsim 11.2$ have approximately constant or mildly increasing radial anisotropy profiles as a function of radius. There is remarkable variety in the sample, with some objects showing a dramatic reduction from radially anisotropic (pressure supported) to tangentially anisotropic (rotation supported) systems with radius, and others instead increasing from tangential anisotropy within \re towards more isotropic values at larger radii. The central values of anisotropy for the sample are also highly diverse. That the orbital structure is not constant with radius is consistent with our understanding of galaxies having internal photometric structures such as bulges and discs, which clearly must reflect different orbital families.

\begin{figure}
\centering
\includegraphics[width=0.95\linewidth]{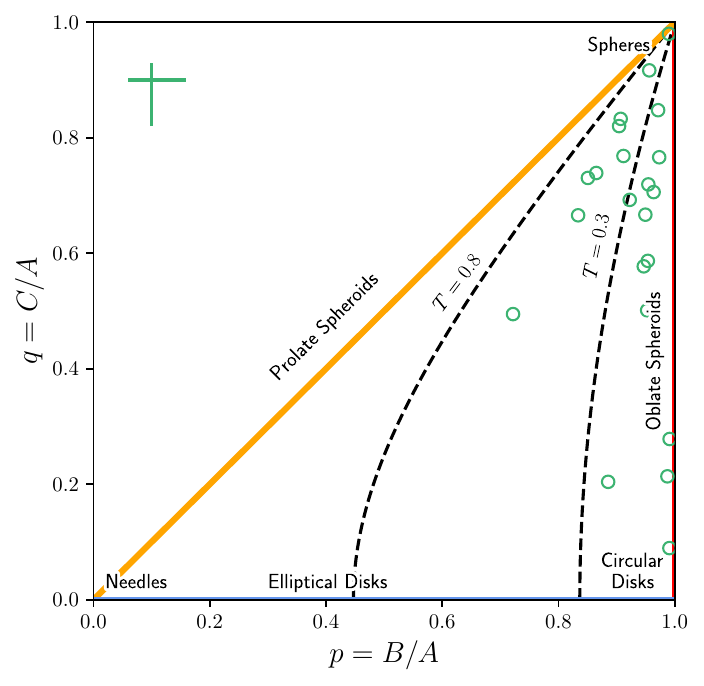}
\caption{The sample in `ellipsoid land', after figure 1 by \citet{de_zeeuw_1991_structure}. This shows the plane of  axial ratios $p = B/A$ and $q = C/A$ and their limiting cases (spheres, needles, and discs). We show curves of constant triaxiality with dashed lines at $\Tr = 0.3$ and $\Tr = 0.8$. Classically oblate spheroids have zero triaxiality. We find few MAGPI objects that populate the strict oblate spheroid zone with $\Tr < 0.1$; most are spheroids but with non-negligible triaxiality. The green dagger in the top left corner shows the median (asymmetric) errors on the intrinsic axial ratios.}
\label{fig:needle_disk_sphere}
\end{figure}

\begin{figure}
\centering
\includegraphics[width=0.6\linewidth]{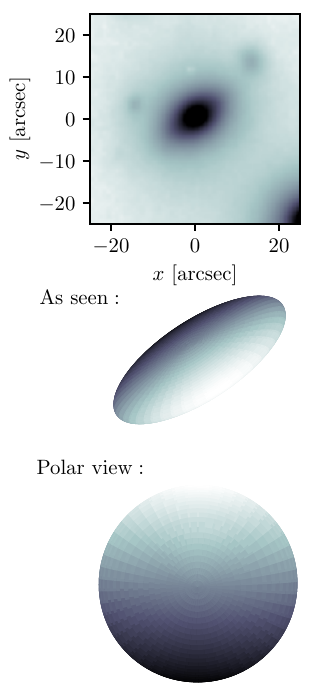}
\caption{The top panel shows the flux image of 152517022 within its MAGPI field. The middle panel shows a rendering of the 3D ellipsoid reconstructed from the Schwarzschild best-fit axial ratios $p = 0.98$ and $q = 0.16$, representing the \textit{intrinsic} shape of the object. The fitted viewing and position angles reproduce the perspective of this object from Earth. The bottom panel shows a $z$-axis view of the intrinsic 3D ellipsoid, to illustrate what a polar view of the object would look like, if only we were able to observationally to access it! This object has negligible triaxiality, and is an oblate spheroid.}
\label{fig:shape_plot:1525170222}
\end{figure}

In Figure~\ref{fig:needle_disk_sphere} we present the intrinsic shapes of the MAGPI sample on the plane of possible ellipsoid shapes in relation to the triaxiality parameter \citep{de_zeeuw_1991_structure}. We find that the majority of the galaxies are consistent with being ellipsoids of varying degrees of triaxiality, with a few tending towards circular (i.e. highly flattened, oblate) discs in their intrinsic shapes. We construct the 3D, ellipsoidal intrinsic shape for MAGPI object 1525170222 in Figure~\ref{fig:shape_plot:1525170222}, which shows how the intrinsic shapes of the best-fit Schwarzschild model translates to an ellipsoid as seen from different viewing angles. Only 3/22 MAGPI galaxies are formally consistent with oblate spheroids (triaxiality parameter $\Tr <0.1$). The remainder show a spread of triaxiality, with 10/22 consistent with mild triaxiality ($0.1 \leq \Tr <0.3$), and 9/22 having values of $\Tr$ spanning $0.3-0.7$. There are no prolate objects.

Unlike what we find for the MAGPI sample, \citet{Santucci_2022} found that 73 per cent of their SAMI sample were oblate spheroids, with only 8 per cent having triaxiality $>0.3$, also inferred from Schwarzschild dynamical models. \citet{emsellem_2011_III} found the vast majority of early-type systems in the \atlas Survey are consistent with being a family of oblate spheroids, with only 12 per cent likely to exhibit mild triaxiality; however, we note this analysis makes inferences from the observed proxy-for-spin parameter and apparent ellipticity rather than dynamical models directly. Additionally, \citet{emsellem_2011_III} did not explore triaxiality as a free parameter, but considered how consistent the assumption of strict axisymmetry was with the data. There is also a mass dependency to the inferred intrinsic shapes from that work, with lower mass systems being more likely to be regular in their rotation with oblate, disc shapes. Similarly, \citet{jin_2020} found a mass dependency with the intrinsic shapes of galaxies, with more massive galaxies more likely to be prolate. 

Some recent results are pointing to galaxies exhibiting triaxial intrinsic shapes rather than axisymmetric ones, in agreement with our MAGPI results. Initial results from Schwarzschild modelling of \atlas galaxies shows that the majority are mildly triaxial or triaxial in shape, with only a few classically oblate systems \citep{Thater_2023_atlas}. Intrinsic shapes were inferred from kinematic alignment and ellipticities for 90 galaxies in the MASSIVE Survey \citep{Ma_2014}, finding most galaxies in the sample are mildly triaxial \citep{ene_2018}.  Complex intrinsic shapes are also observed in  the Magneticum simulations, with \citet{valenzuela_2024} finding most galaxies are triaxial. However, that work also found a high precedence of prolate galaxies, of which we have none in the presented MAGPI sample. 

\subsection{Systematics between Schwarzschild and Jeans modelling}
\label{sec:results:jeans}

\begin{figure}
\centering
\includegraphics[width=\linewidth]{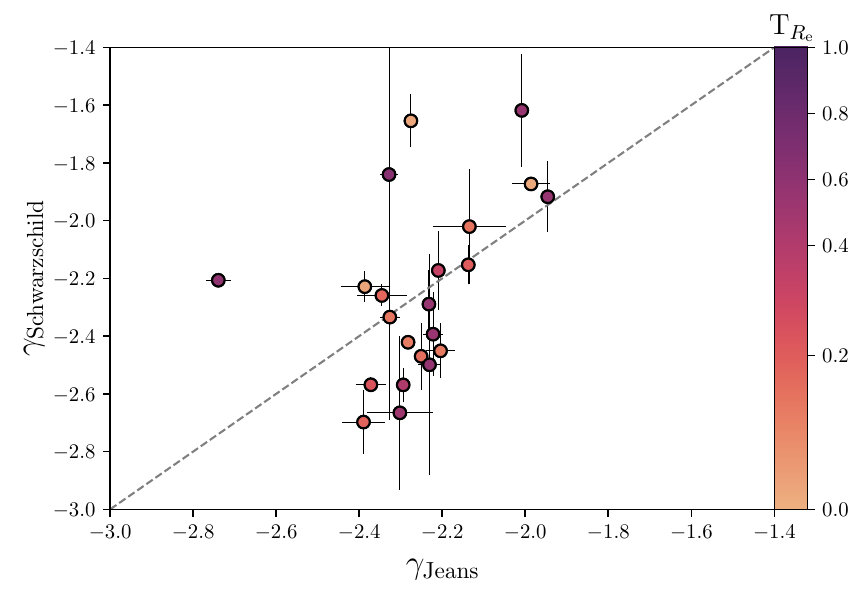}
\caption{A comparison between the total density slopes derived using Jeans models and presented in \citet{derkenne_2023_MAGPI} and those derived using Schwarzschild models in this work. The 1-1 line is shown. The galaxies are coloured by triaxiality, \tr. }
\label{fig:systemmatics}
\end{figure}

The MAGPI galaxy sample presented here is a subset of the 30 galaxies with total density slopes measured using anisotropic Jeans models and a power-law potential by \citet{derkenne_2023_MAGPI}, which allows for an examination of the systematics between the two modelling methods on the same objects. Jeans dynamical modelling, under the assumption of a power law total density profile, is a powerful method especially when considering objects at significant redshift. When deriving the total density profile of a galaxy, the absolute calibration of the surface brightness is unimportant; only the relative scaling between MGE components matters to act as the luminous density tracer for the stellar kinematics. As such, assuming a total potential means the models are agnostic to many redshift effects and sources of systematic errors, such as cosmological surface brightness dimming, and how stellar light corresponds to stellar mass. Alternatively, building the gravitational potential from the stellar luminosity density and a separate dark matter halo, as presented in this work, requires more careful calibration of the stellar components (and assumptions around the IMF). Details of the Jeans modelling method can be found in \citet{cappellari_measuring_2008}, with specific details of how the MAGPI sample was modelled with the Jeans method in \citet{derkenne_2023_MAGPI}. In Figure~\ref{fig:systemmatics} we show the total density slope comparison of the Jeans and Schwarzschild models, with objects colored by their triaxiality. 

We note a few differences between the modelling approaches. In the Jeans modelling approach used in \citet{derkenne_2023_MAGPI}, we assumed axis-symmetry, a global but free orbital anisotropy, a cylindrically aligned velocity ellipsoid, and NFW-like double-power law with a free inner slope. This power-law forms the \textit{total} gravitational potential, in that we made no attempt to disentangle the baryonic and dark components. In the Schwarzschild method used in this work, the models are fully triaxial and orbital structures can vary freely due to the three-integral orbit-based superposition method of the models. Furthermore, instead of assuming a total density profile shape, we assume the dark matter profile and measured the stellar mass profiles from a combination of photometry and spectral stellar population mass-to-light ratios. Due to this method, it is possible for the fitted total mass density profile to deviate from a power law.

\begin{figure}
\centering
\includegraphics[width=\linewidth]{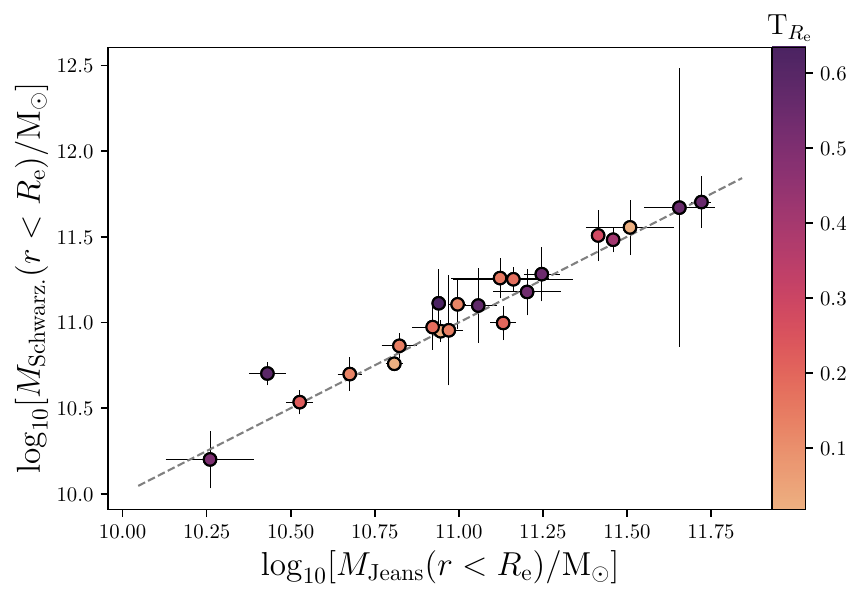}
\caption{A comparison between the dynamical masses measured using the Schwarzschild (this work) and Jeans \citep{derkenne_2023_MAGPI} methods. The 1-1 line is shown. The galaxies are coloured by triaxiality, \tr. }
\label{fig:mass_agreement}
\end{figure}

From the direct comparison of these two methods, we can see that in general rank is preserved: steep slopes as measured by Jeans models remain steep when measured using Schwarzschild models, and shallow slopes remain comparatively shallow (with a notable outlier). However, the slopes generally do not agree within the errors. The discrepancy cannot be explained by projection effects, as there is no trend between the triaxiality of the sample and the offsets from the 1-1 line. There is only a mild (and non-significant) correlation between the slopes measured with different methods, with a Pearson correlation coefficient of 0.42 and p-value of 0.05. By contrast, the dynamical masses with an effective half-light radius are well recovered, with almost all the dynamical masses agreeing within the estimated uncertainties, as shown in Figure~\ref{fig:mass_agreement}. Despite a direct galaxy-by-galaxy comparison being inconsistent, the median total density slopes measured by the two methods agree well. From the Schwarzschild models in this work we obtain a median density slope of $\gamma_{\mathrm{tot}} = -2.27 \pm 0.08$, and for the overlapping Jeans sample we find $\gamma_{\mathrm{tot}} = -2.26 \pm 0.04$. 

As argued by \citet{etherington_2023_beyond} when comparing a lensing-only analysis of density slopes with a lensing and dynamics approach, a 1-1 correspondence should be expected if the true density profile of the systems are well represented by a power law. That the slopes do not agree, for the most part, within the errors, is evidence that the MAGPI systems deviate from strict power law potentials. The extra freedom of the Schwarzschild models, in terms of both modelling assumptions and the gravitational potential, yield different slopes to the Jeans derived ones. Interestingly, the variance of the Jeans derived slopes are much lower than that of the Schwarzschild models. The Schwarzschild sample total slopes have a standard deviation of $0.30 \pm 0.03$, whereas the Jeans total slopes have a standard deviation of $0.16 \pm 0.03$. We suggest this is because the total density  profile shape is assumed to be a power law in the Jeans models, but for the Schwarzschild models the total profile is some free combination of the stellar mass profile and a dark matter halo. The additional freedom allowed by the gravitational potential used for the Schwarzschild models permits a larger radial variation of intrinsic density profiles to be fitted. Assuming this diversity is present in real galaxies, this naturally gives rise to larger galaxy-to-galaxy variation in the inferred density slopes.

Recently, a sample of over 10,000 MaNGA galaxies was dynamically modelled by \citet{zhu_2023_dynpop1} using multiple Jeans method approaches, all of which had a gravitational potential constructed by including the stellar luminosity information with the possible co-addition of different dark halo profile types (in a similar vein to the approach used in this work). \citet{zhu_2023_dynpop1} found that the total slopes, measured using the different assumed dark matter haloes, were consistent. However, for the Jeans dynamical models constructed by \citet{poci_2017_syst} on the \atlas sample, the mean total slope as measured by assuming a power law total density profile structure was not consistent with the total slope as measured by building the density profile from the de-projected stellar luminosity and an assumed dark halo (comparing the Model I and Model III total density profile slope means in that work). 

From the above, we suggest it is not the dynamical modelling method (Jeans or Schwarzschild) \textit{per se} that causes the systematic difference observed here, but rather the assumption of the structure of the total gravitational potential. It may also be that in assuming a  power law density profile structure when measuring total density slopes, as is done in various dynamical and lensing approaches, artificially reduces the population variance observed. A thorough exploration of the systematics between the modelling techniques and assumptions would be required to fully understand this issue, and is beyond the scope of this work.

\section{Discussion}
\label{sec:discussion}

The primary goal of this work is to use the high quality kinematic data from the MAGPI Survey to investigate the stellar and total density profiles of massive galaxies using a highly general dynamical modelling approach at a cosmologically significant look-back time. We have shown no correspondence between central dark matter fractions and stellar density profiles, whilst dark matter is highly correlated to the gradient of the \textit{total} density profile. These results are in contrast gravitational lensing studies, which consistently report low intrinsic scatter in measured total density slope values, pointing to a homogeneity for (early-type) galaxy mass density structures. 

A limitation of our approach has been to use a $M_{200}-c$ coupled NFW halo, which effectively means that the only freedom in the total density profile is the combination of a globally scaled stellar mass profile and the dark matter fraction. As seen with some galaxies in the sample, it is possible to reproduce the higher order kinematics with only the stellar profile and negligible contributions from the dark halo within the data range probed. Because we have not implemented the effects of IMF, if the true total density profile is even steeper than the measured stellar profiles, our models are bounded at having no dark matter. As a result, it is possible that constraints caused by our modelling approach has impacted the resulting total density slopes returned. In our comparison of the Schwarzschild models against Jeans dynamical models of the same sample in Section \ref{sec:results:jeans}, we show that our Schwarzschild model slopes are more variable than those returned using Jeans models with a completely free power-law slope on the total density profile and no assumption on the form of the stellar or dark matter profiles. This suggests our Schwarzschild models are not too restrictive in their assumptions. Furthermore, if our models lacked significant freedom we would expect to see departures of the modelled kinematics to the observed ones. Instead, we note our models are able to reproduce the higher order kinematic features of many galaxies in the sample, such as MAGPI object 1530197196. This aside, a modelling approach with, for example, a generalised NFW dark matter halo with free inner slope and the inclusion of variable IMF would be invaluable in testing the bulge-halo conspiracy in the future.

A recent meta-analysis of gravitational lensing works  by \citet{etherington_2023_beyond} examined 48 lensing systems from the Sloan Lens ACS \citep{bolton_2006_sloan,auger_sloan_2010}, BOSS Emission Line Lensing \citep{Brownstein_2012}, BELLS GaLaxy-Ly$\alpha$ Emitter Systems \citep{shu_2016_gallery}, Strong Lensing Legacy Survey \citep{gavassi_2012_sl2s}, and the Lenses, Structure and Dynamics \citep{Treu_2004} surveys, with total density slopes calculated with two different methods: using a lensing-only approach as presented in \citet{etherington_2022_no}, and a joint lensing and dynamics approach, where the stellar velocity dispersion is used to constrain the mass density profile. Consistent with the bulge-halo conspiracy, \citet{etherington_2023_beyond} found a mean, almost isothermal, total density slope for the sample of $\gamma_{\mathrm{tot}} = -2.075^{+0.023}_{-0.024}$ with an intrinsic scatter of $\sigma_{\gamma} = 0.172^{+0.022}_{-0.032}$, assuming a power-law density distribution. This scatter is far smaller than what we find for the MAGPI sample, of $\sigma_{\gamma} = 0.30 \pm 0.03$.

It is unclear what drives the differences between the variation in total slope values reported in this work and those found by lensing studies, although it possibly driven by how the gravitational potential is incorporated in the lensing models compared to the more general potential we have allowed for in this work. Furthermore, \citet{etherington_2023_beyond} note that lensing-only analyses are sensitive to the local slope at the Einstein radius, which is more likely to occur at a transitional region between the stellar dominated centre and dark matter dominated outskirts, and so lensing derived slopes are more likely to indicate possible deviations from a power-law density structures for galaxies. Lensing in combination with dynamical methods tends to average the slope between the inner and outer galaxy regions, potentially fueling the conspiracy, as deviations from a  power-law structure may be averaged out. Methodologically this might resolve the conspiracy as lensing-only studies are conducted across larger redshift baselines.

The bulge-halo conspiracy has also been explored with 5688 Jeans dynamical models from local Universe MaNGA galaxies \citep{dynapop_bulge_halo} with visual checks performed to ensure model quality. Specifically, \citet{dynapop_bulge_halo} explored the correlation between the stellar density slopes, total slopes, and dark matter fraction within the effective radius of their sample. As found in this work, they found a steep stellar density slope corresponds to a steep total density slope, whereas a high dark matter fraction corresponds to a more shallow total density slope (see their figures 6 and 7). \citet{dynapop_bulge_halo} found no evidence of the conspiracy for their full sample, in that there is a wide range of slopes observed (between $-1.2$ and $-2.8$). For early type galaxies with high velocity dispersions, \citet{dynapop_bulge_halo} found the median total density slope does trend towards near isothermal values, but that there is still a correspondence between stellar density profiles and total density profiles. That is, dark matter does not compensate to create a conspiracy. Instead, \citet{dynapop_bulge_halo} conclude that the outcome of a median isothermal total density slope for early types is simply a coincidence resulting from the mass growth of galaxies.

In an accompanying work to \citet{dynapop_bulge_halo}, \citet{dynapop_stellar_pops} found that galaxies with older stellar populations tend to have steeper total density slopes, and galaxies with young stellar populations have shallow total density slopes (in particular, see their figure 8). This trend is also coupled to the central velocity dispersions, such that galaxies with high velocity dispersions have comparatively steeper slopes than galaxies with lower velocity dispersions, as initially seen by \citet{poci_2017_syst}. \citet{dynapop_stellar_pops} further find that the scatter in total density slopes also decreases with increasing central velocity dispersion. The MAGPI galaxies typically have high velocity dispersions, with a range of $97 - 328\,\,\mathrm{kms}^{-1}$ and a mean of $210\,\,\mathrm{kms}^{-1}$, which is on the high end of the MaNGA sample distribution. Even with a high average velocity dispersion, we still find large scatter in the MAGPI total density slopes.

In exploring the conspiracy with simulation data, \citet{remus_2013_dark} found a power-law approximation to be a reasonable fit for total mass density profiles (see their figure 2). In the simulations used, merger events caused the total density slope to flatten towards isothermal values, with steep total density slopes likely for systems that have undergone few mergers and therefore have a high fraction of in situ stars. Elliptical galaxies with almost no mergers in their history  tended to retain steep density slopes of $\gamma \sim -3$, whilst progressive dry mergers cause the slope to converge nearer to $\gamma \sim -2$ across cosmic time, coupled with increasing dark matter fractions \citep[see][in particular, their figure 12]{remus_2013_dark}.

Minor mergers in particular are efficient in building galaxies with high dark matter fractions, as their stars are incorporated into the main progenitor at relatively large radii where dark matter is more dominant, while also reproducing the observed size growth for both elliptical galaxies \citep{hilz_2012_relaxation,hilz_2013_minor} and disc galaxies \citep{karademir_2019_outer}.  In this sense, an isothermal total density  slope of $-2$  might act as a natural attractor due to progressive mergers, as noted by \citet{rhea_2017_coevolution}, although values both considerably steeper and more shallow than this are found in both simulations and observations \citep{rhea_2017_coevolution,poci_2017_syst,bellstedt_sluggs_2018,derkenne_2023_MAGPI,zhu_2024_manga}. Due to the relation between dark matter fractions, mergers, and in situ stars, \citet{rhea_2017_coevolution} also found an anti-correlation between the dark matter fractions and fraction of in situ stars with the Magneticum simulations.  Instead of a conspiracy, the evolution of the galaxy's total density profile towards shallower values, from steep values in  the early Universe, indicates the (stochastic) merger history of a galaxy. Following this line of argument, we would naturally expect the variation of total density profiles we see with MAGPI galaxies, instead of a homogeneous sample of near-isothermal total density slopes. As gas accretion becomes less common at more recent times \citep{deRavel_2009_VIMOS,Oser_2010,tacconi_2010_high}, the theorised evolution is for total density slopes to become more shallow, although we note the reverse of this evolution is seen by lensing samples \citep{etherington_2023_beyond}.  

Finally, there are clearly systematic effects at play between building the gravitational potential under the assumption of a power law total density profile with Jeans dynamical modelling, and a Schwarzschild modelling approach with a potential constructed from the combination of a stellar mass profile and separate dark matter halo, as shown in Section~\ref{sec:results:jeans}. These systematics are despite the fact both methods use almost identical data inputs (MGEs and 2D stellar kinematics). We can only assume that there are further systematic effects to be considered between dynamical modelling in general and lensing, and so advocate as future work the joint analysis of the same systems with the available, complementary, techniques, as done by \citet{poci_2022}, but extended to larger samples of galaxies. Only in this way will we ensure that comparisons across studies, and the conclusions drawn from them, are fairly made. 

\section{Summary and Conclusions}
\label{sec:conclusion}
We present Schwarzschild orbit-based triaxial models for a sample of \Nsample well-resolved galaxies from the MAGPI Survey, exploring in detail the properties of massive galaxies during the Universe's middle ages. The sample spans stellar masses between $\log_{10}(M_{\star}/\Msun) = 10.4$ and $\log_{10}(M_{\star}/\Msun) = 11.6$, with a median stellar mass of $\log_{10}(M_{\star}/\Msun) = 11.2$ The MAGPI Survey uses MUSE/VLT in the wide-field mode coupled with ground layer-adaptive optics, resulting in rich and highly resolved fields  at a target redshift of $z \sim 0.3$. We use 2D, 4-moment stellar kinematics from full spectrum fitting and multi-Gaussian expansions of the stellar light, in combination with spectral mass-to-light ratio maps, to construct the Schwarzschild models.  We use the state of the art {\sc{dynamite}} code \citep{jethwa_2020_dynamite}, allowing for fully triaxial Schwarzschild models, and make no assumptions on orbital structure. To our knowledge, this is the first time Schwarzschild modelling has been applied on a sample of galaxies at a cosmologically significant redshift. 

Instead of assuming a power law for the gravitational potential, as commonly done in both lensing and dynamical studies of galaxies, we instead measure the stellar surface brightness profile and transform it to a stellar mass profile using spectral stellar mass-to-light ratio maps from MAGPI data. To build the potential we combine the stellar mass profile with a freely scaled NFW dark matter halo, in order to test the so-called bulge-halo conspiracy. Numerous observations have recovered near isothermal total (baryonic and dark) density profiles for galaxy populations, with small intrinsic scatter. This finding requires that the dark and luminous mass components must somehow conspire to yield remarkable homogeneity in observed galaxy mass structures despite varied stellar light structures. From our models we obtain both stellar mass profiles and constrained NFW profiles. In addressing the bulge-halo conspiracy, we find:

\begin{enumerate}

    \item The bulge-halo conspiracy predicts that the variety of structures observed for the stellar component and theorised for the dark component are larger than the variation of the combined, approximately isothermal, profile. Fitting a flexible potential with separate realistic stellar and dark mass profiles, we find a wide range of total density profile slopes, spanning approximately between $\gamma = -2.7$ and $\gamma = -1.6$ for $\rho \propto r^{\gamma}$ and $\gamma < 0$. Indeed, the dispersion of the total density profile slopes  is greater than that for the stellar profile slopes alone, comparing $\sigma_{\mathrm{tot}} = 0.30 \pm 0.03$ to $\sigma_{\star} = 0.19 \pm 0.02$ for the total and stellar components, respectively. This finding is inconsistent with the bulge-halo conspiracy. 
    \item While we find a tight correlation (Pearson $r > 0.8$) between the total density profile slope and the central dark matter fraction, and a moderate correlation between the total density profile slope and the stellar profile slope, we find no significant correlation between the central dark matter fraction and the stellar profile slope.  The lack of correlation between the stellar mass structure and the dark matter fraction is also inconsistent with the bulge-halo conspiracy, as the dark matter should systematically adjust with differing stellar profiles if the conspiracy were valid. However, as predicted by the Magneticum simulations, MAGPI galaxies with  higher central dark matter fractions have shallower total density profiles. The relation between the dark matter fraction within the half-light radius and density slopes is $\gamma_{\mathrm{tot}} = (1.3 \pm 0.2) f_{\mathrm{DM}} - (2.44 \pm 0.04)$, consistent with Magneticum results \citep{rhea_2017_coevolution}.  
    \item From a comparison of total mass density profile slopes from the  Schwarzschild and Jeans  modelling methods, we suggest that assuming a power law density profile may artificially reduce the observed intrinsic scatter in population density profile slopes. 
\end{enumerate}

Our detailed Schwarzschild models also allow us to explore other intrinsic properties of the sample, including orbit types and structures, and the intrinsic shapes of massive galaxies. In exploring the MAGPI sample with these models, we find:
\begin{enumerate}
    \item The median 1 \re dark matter fraction for the sample is 10 per cent with a standard deviation of 19 per cent, which is similar to local Universe survey results from MaNGA, but lower than reported from the SAMI Survey.
    \item The dark matter profile does not tend to dominate at any specific radius across the sample. For some galaxies the dark matter density exceeds the stellar density within the half-light radius, while for others the dark matter only exceeds the stellar mass density at more than 10 \re. How baryonic and dark matter is distributed within a galaxy is therefore highly variable. 
    \item In exploring the intrinsic shapes of our MAGPI sample at 1 \re, we find that only 3/22 are formally consistent with oblate spheroids (triaxiality parameter $\Tr <0.1$). The remainder show a spread of triaxiality, with 10/22 consistent with mild triaxiality ($0.1 \leq \Tr <0.3$), and 9/22 having values of $\Tr$ spanning $0.3-0.7$.
    \item The triaxiality of a galaxy correlates strongly with the fraction of hot orbits within an effective radius, as well as the orbital anisotropy of the galaxy. Galaxies that exhibit radial anisotropy are likely to be strongly triaxial in intrinsic shape. 
    \item The anisotropy profiles of the sample are also highly variable. Massive galaxies with radial anisotropy tend to have constant or mildly increasing anisotropy with radius, whereas less massive galaxies that are rotation supported can display either increasing or decreasing tangential anisotropy with radius. The central anisotropy within the half-light radius is also variable for the sample.  
\end{enumerate}

From our results, we conclude that the internal mass structure of galaxies tends to be non-homogeneous, varying both between galaxies and within galaxies. Our analysis acts as a caution in making strong assumptions when constructing dynamical (or lensing) models of galaxy internal mass distributions and structure. The more closely we examine galaxies, the more rich in structure they reveal themselves to be - and with the increasing observing power of modern integral field units, we are continuously improving our ability to explore distant galaxies in unprecedented detail. 

The detailed orbital structures revealed in this analysis of $z \sim 0.3$ galaxies is only with a sub-sample of MAGPI Survey targets, with tens more objects with sufficient data quality for such modelling expected from the completed survey. The advance of both space-based and ground-based instruments, for example JWST and extremely large telescopes such as the E-ELT, mean that it is possible for this kind of analysis to be pushed yet further afield, connecting the local Universe to its middle ages and beyond. The discovery of $z \sim 2$ lenses presents the opportunity of joint lensing and dynamics studies at lookback times of 10 Gyr \citep[e.g.][]{Mercier_2023_cosmos,dokkum_2024}. The instrumentation available is starting to allow for consistently applied methods across unprecedented cosmic time baselines.  

\section*{Acknowledgements}
We warmly thank the reviewer for their comments which helped improve this work.
We wish to thank the ESO staff, and in particular the staff at Paranal Observatory, for carrying out the MAGPI observations. MAGPI targets were selected from GAMA. GAMA is a joint
European-Australasian project based around a spectroscopic campaign using the Anglo-Australian
Telescope. GAMA was funded by the STFC (UK), the ARC (Australia), the AAO, and the participating
institutions. GAMA photometry is based on observations made with ESO Telescopes at the La Silla
Paranal Observatory under programme ID 179.A-2004, ID 177.A-3016. The MAGPI team acknowledge support by the Australian Research Council Centre of Excellence for All Sky
Astrophysics in 3 Dimensions (ASTRO 3D), through project number CE170100013. SMS acknowledges funding from the Australian Research Council (DE220100003).
CF is the recipient of an Australian Research Council Future Fellowship (project number FT210100168) funded by the Australian Government.
CL, JTM and CF are the recipients of ARC Discovery Project DP210101945.

\section*{Data Availability}

The MAGPI spectral data underlying this work are available from the ESO Science Archive Facility: {\url{http://archive.eso.org/cms.html}}.

\section*{Software Citations}
This work uses the following software packages: 

\begin{itemize}
    \item  \href{https://github.com/astropy/astropy}{\sc{astropy}} \citep{astropy_collaboration}
    \item \href{https://github.com/dynamics-of-stellar-systems/dynamite}{\sc{dynamite}} \citep{jethwa_2020_dynamite}
    \item \href{https://users.obs.carnegiescience.edu/peng/work/galfit/galfit.html} {\sc{galfit}} \citep{Peng_2002_galfit}
    \item \href{https://pypi.org/project/ltsfit/}{\sc{ltsfit}} \citep{cappellari_2013_benchmark}
    \item \href{https://github.com/matplotlib/matplotlib} {\sc{matplotlib}} \citep{Hunter:2007}
    \item \href{https://pypi.org/project/mgefit/}{\sc{mgefit}} \citep{MGE}
    \item \href{https://github.com/numpy/numpy}{\sc{numpy}} \citep{harris2020array}
    \item \href{https://pypi.org/project/pafit/}{\sc{pafit}} \citep{krajnovic_2006_kinemtry}
    \item \href{https://github.com/pandas-dev}{\sc{pandas}} \citep{mckinney-proc-scipy-2010}
    \item \href{https://pypi.org/project/plotbin/}{\sc{plotbin}}
    \item \href{https://pypi.org/project/ppxf/}{\sc{ppxf}} \citep{cappellari_measuring_2008,cappellari_2020}  
    \item \href{https://github.com/asgr/ProFound}{\sc{profound}} \citep{robotham_2018_profound}
    \item \href{https://github.com/asgr/ProSpect}{\sc{prospect}} \citep{robotham_2020_prospect}
    \item \href{https://pypi.org/project/vorbin/}{\sc{vorbin}} \citep{cappellari_adaptive_2003}
    \item \href{https://github.com/scipy/scipy}{\sc{scipy}} \citep{2020SciPy-NMeth}
    \item \href{https://github.com/musevlt/zap}{\sc{zap}} \citep{soto_2016_zap}
 
\end{itemize}



\bibliographystyle{mnras}
\bibliography{schwarzschild}



\appendix

\section{Impact of dithering on model orbit weights}
\label{sec:appendix:dithering}
\begin{figure}
\centering
\includegraphics[width=\linewidth]{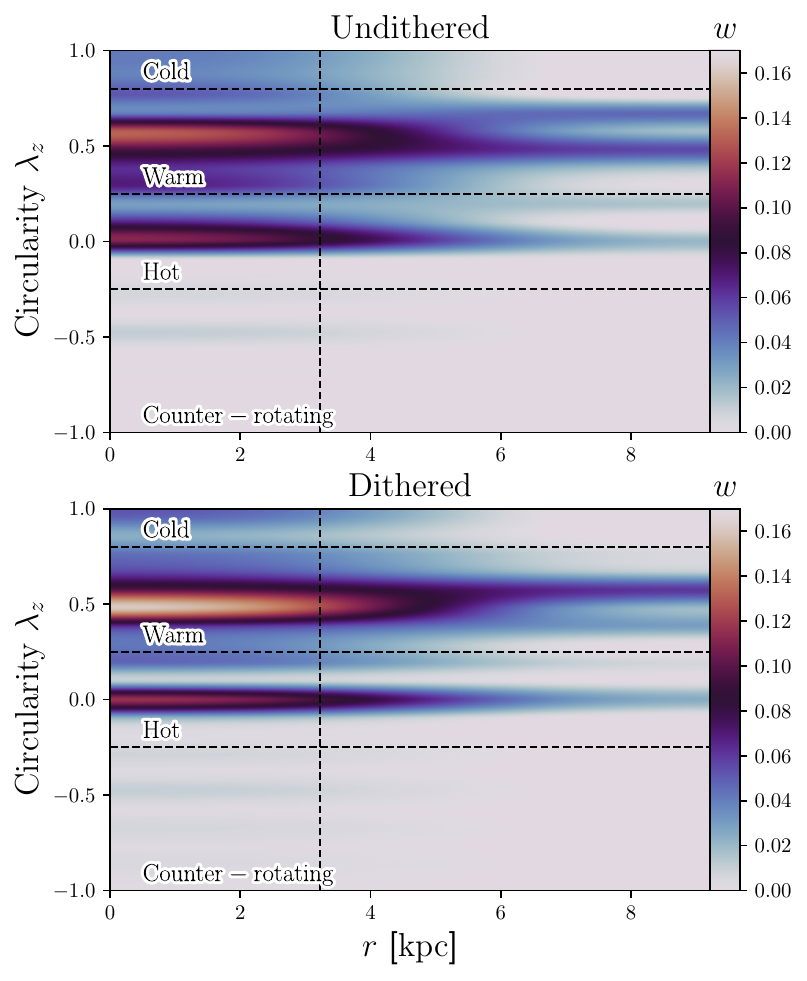}
\caption{The orbital weights, $w$, for MAGPI object 1203060081, with no dithering (top) and dithering set to 3 (bottom). The dashed-line shows one effective radius, and the x-axis extends to the extent of the kinematic data.}
\label{fig:dither_v_undither}
\end{figure}

We chose not to use dithering for our Schwarzschild models. Dithering perturbs the initial starting positions of the orbits in the (E, $L_z$, $I_3$) space to regularise the final weighted orbits. Dithering increases the orbit library size by a factor of $N^{3}$, where N is the chosen dithering of, for example, 3 or 5. For a technical description and motivation for dithering see \citet{bosch_2008_triaxial}. Due to the increased orbit library size, dithered models take significantly more computational time to run than undithered models. We explored whether dithering made a substantial impact on the returned model orbit weights by running a test MAGPI galaxy (ID = 1203060081) without dithering, and then with dithering set to 3. We found that there was no significant change in the orbit weights for the test galaxy, however the undithered model took of order ~400 hours to run, whereas the dithered model took ~8000 hours of equivalent compute time. We show the orbital weights for the dithered and undithered best-fit models in Figure \ref{fig:dither_v_undither}. We conclude that for MAGPI-like data quality, dithering is not necessary. 

\section{Sample Schwarzschild model kinematic moments}
\label{sec:appendix:models}

In Figure~\ref{fig:first_eight} we show the remaining 20 Schwarzschild models for the MAGPI sample used in this work, with models for MAGPI objects 1205093221 and 1525170222 shown in Figures~\ref{fig:1205093221_kinematics} and \ref{fig:1525170222_kinematics}, respectively. For each model we show show the surface brightness and line-of-sight kinematic moments up to $h_4$. The panels are ordered by MAGPI field number, being the first four digits of the MAGPI ID. Gaps in the fields are for masked regions, e.g. for MAGPI object 1208197197.

\begin{figure*}
\centering
\includegraphics[width=\linewidth]{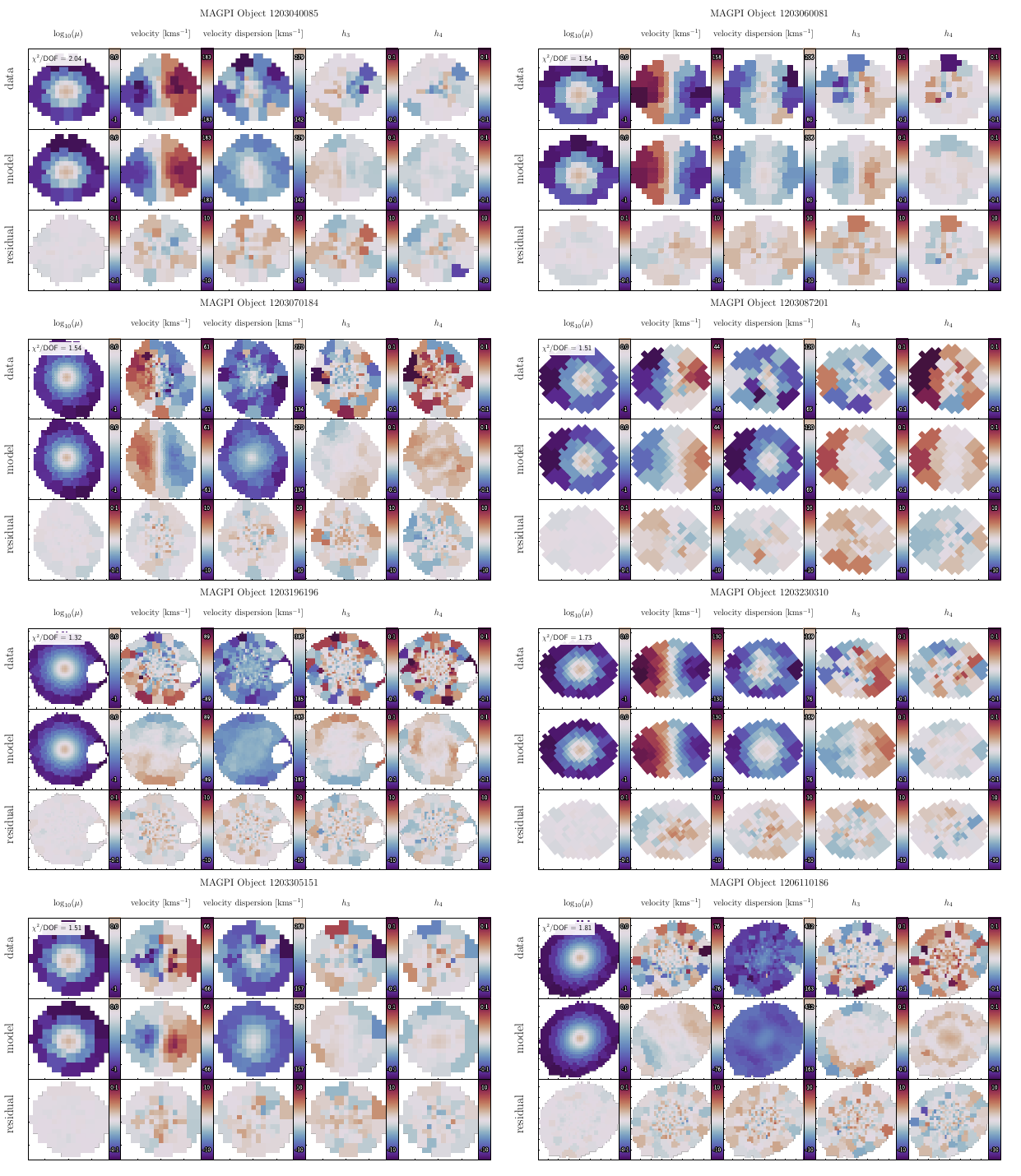}
\caption{The Schwarzschild models for the remaining 20 MAGPI objects not show in the main text, with the figure details the same as for Figure~\ref{fig:1205093221_kinematics}.}
\label{fig:first_eight}
\end{figure*}

\begin{figure*}
\centering
\includegraphics[width=\linewidth]{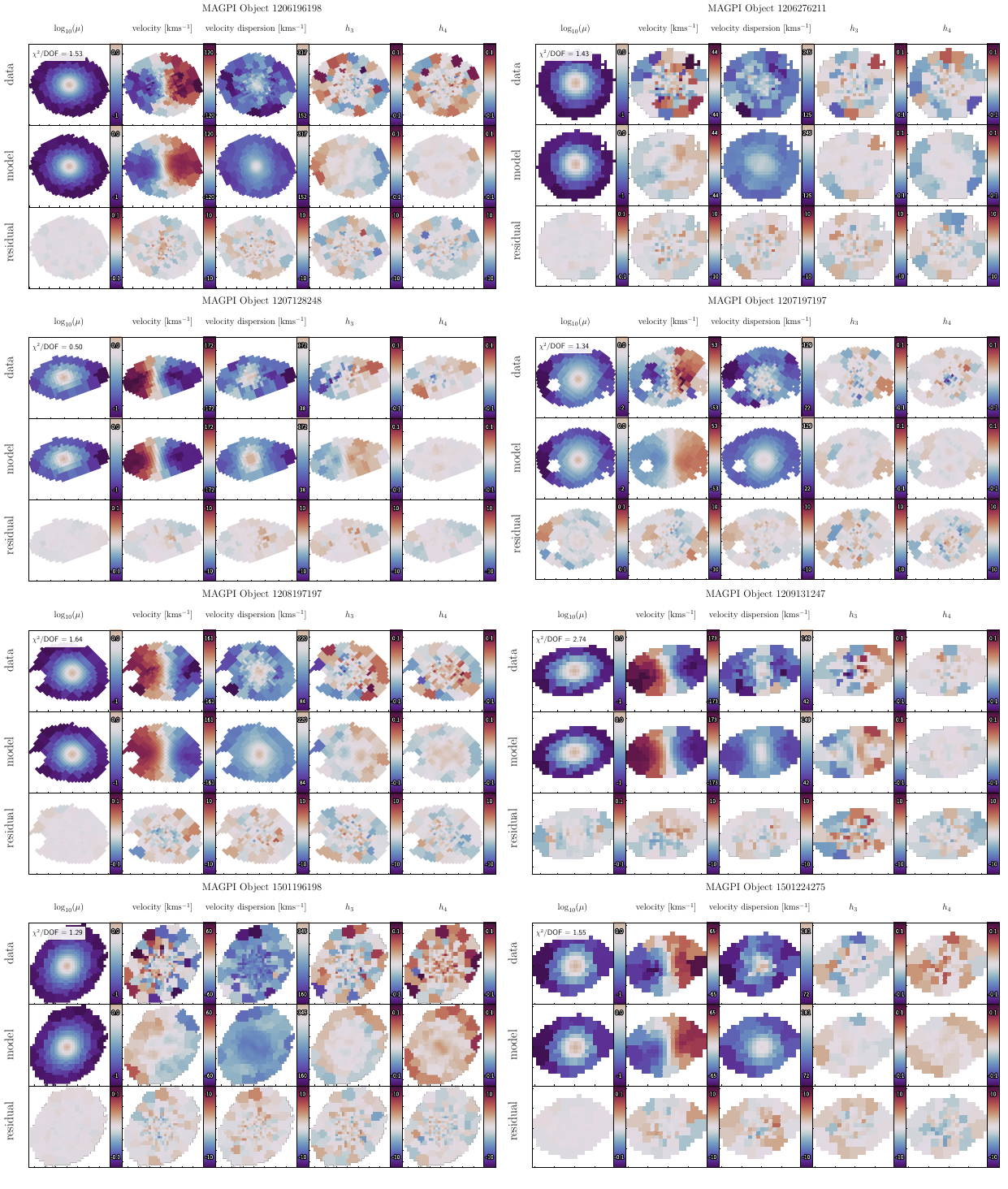}
\contcaption{}
\label{fig:second_eight}
\end{figure*}

\begin{figure*}
\centering
\includegraphics[width=\linewidth]{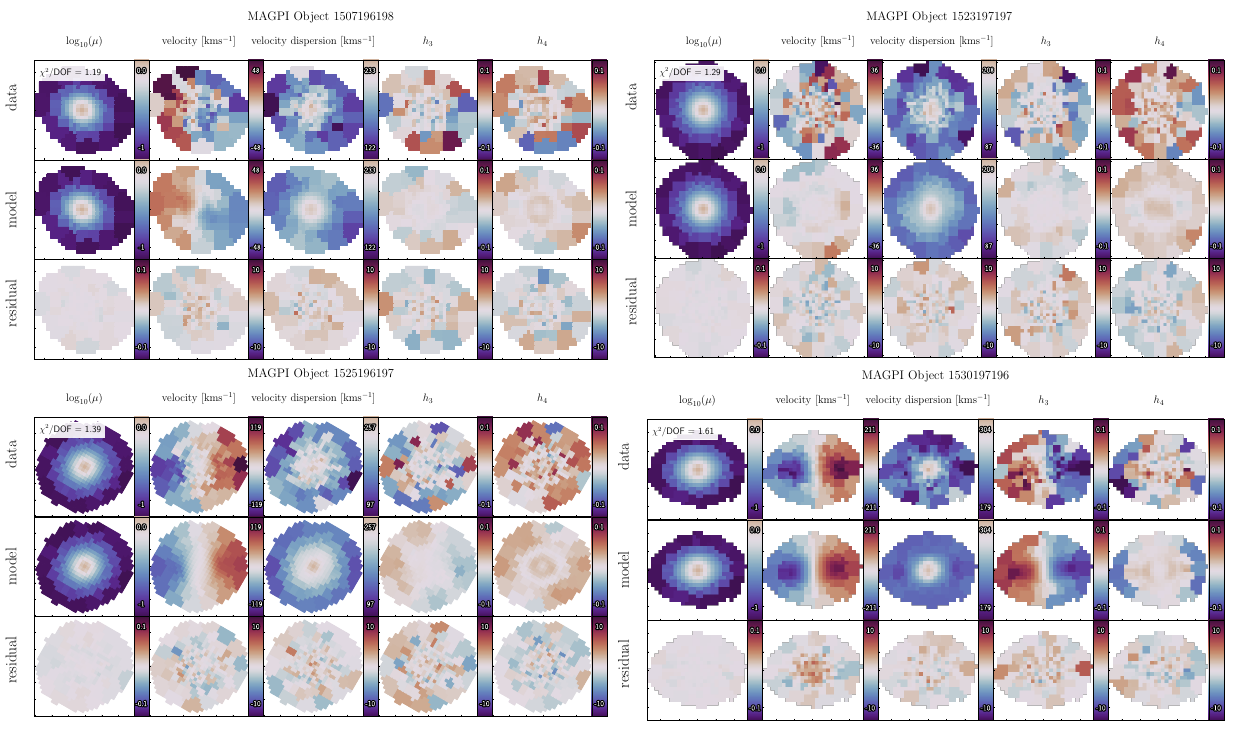}
\contcaption{}
\label{fig:last_four}
\end{figure*}

\section{Testing the impact of variable IMFs}
\label{sec:appendix:imf}

In our modelling methods we have so far left the thorny question of variable IMFs unanswered. This is partly because the data quality and spectral lines required (e.g. TiO2 and  Na I) to effectively measure the IMF \citep{spiniello_2010_evidence} are unavailable with MAGPI data due to redshift and signal-to-noise. However, we still made an attempt to understand what impact a variable IMF within a galaxy might have on our results. A non-universal IMF \textit{between} objects is not a concern because of the free re-scaling of all the mass components in the Schwarzschild models, parameterised as $\Upsilon_r$, which essentially adjusts the global IMF. Therefore, although all spectroscopic mass-to-light ratios are calibrated to a Salpeter IMF, the mass scaling in the Schwarzschild models is left free and not tied to any particular IMF.

Although a difference of IMF between galaxies is not a concern, an additional gradient on the stellar mass-to-light ratios due IMF structures \textit{within} galaxies could have an effect on the returned dark matter fractions. Our models attempt to accurately reconstruct the stellar mass profile by use of detailed surface brightness models and stellar population mass-to-light ratios, fitting for age and metallicity variations (which are constrained by the MAGPI data). The mass-to-light ratios adjust the gradient of the stellar surface brightness, which in turn  de-distributes the stellar mass. A radially-varying IMF would introduce another gradient to the stellar mass profile.

Observations support a non-global galaxy IMF, in particular the 2D resolved view of Fornax3D \citep{sarzi_2022_fornax3D} galaxy FCC 167 showing a dwarf-rich population in the galaxy's center, which correspond to higher mass-to-light ratios in the central regions compared to the outskirts \citep{navarro_2019_fornax}. A wealth of other studies support this view of IMF gradients \citep{martin_navarro_2014_radial,Conroy_2017,dokkum_2017_IMF}. If we assume that in general massive galaxies tend to be dwarf rich in their centres, then a variable IMF would tend to make the stellar mass profiles steeper. To see what effect this has on Schwarzschild-derived orbits and dark matter fractions, we construct a simple test case. For MAGPI object 1525170222 we artificially steepen the stellar mass profile by increasing the innermost mass-to-light ratio by 15 per cent and decreasing the outermost by the same, and re-run the Schwarzschild models. This adjustment level was chosen because the fraction of mass contained in the dwarf-population FCC 167 changes from 0.66 to 0.5 from the galaxy centre to the outskirts.

For MAGPI object 1525170222 we find the dark matter fraction increases for our test case, to $0.50^{+0.11}_{-0.16}$ from an original $0.44^{+0.11}_{-0.06}$. The global mass scaling $\Upsilon_r$ parameter correspondingly decreased to $0.47^{+0.08}_{-0.08}$ from $0.53^{+0.08}_{-0.04}$. The reduced $\chi^2$ value of the IMF test case and the original model are comparable: 2.45 and 2.43, respectively.  Although this is a single test case, we use this example to illustrate the possible uncertainties, of order 10 per cent, on our reported dark matter fractions due to possible IMF gradients.


\bsp	
\label{lastpage}
\end{document}